\documentclass[acmlarge,screen]{acmart}
\AtBeginDocument{%
  }

\setcopyright{acmlicensed}
\copyrightyear{2018}
\acmYear{2018}
\acmDOI{XXXXXXX.XXXXXXX}

\usepackage{graphicx}
\usepackage{wrapfig}
\usepackage{bbm}
\usepackage{url}
\usepackage{algorithm}
\usepackage{textcomp}
\usepackage{amsmath}
\usepackage[caption=false,farskip=0pt,labelfont={bf}]{subfig}
\usepackage[T1]{fontenc}
\usepackage{booktabs} 
\usepackage{xcolor}
\usepackage{multirow}
\usepackage{color}
\usepackage{wrapfig}
\usepackage{booktabs} 
\begin{document}

\title{AdaptiveLog: An Adaptive Log Analysis Framework with the Collaboration of Large and Small Language Model}

\author{Lipeng Ma}
\email{lpma21@m.fudan.edu.cn}
\orcid{0000-0001-5974-5988}
\affiliation{%
  \institution{Shanghai Key Laboratory of Data Science, School of Computer Science, Fudan University}
  \city{Shanghai}
  \country{China}
}

\author{Weidong Yang}
\authornote{Corresponding authors.}
\orcid{0000-0002-6473-9272}
\affiliation{%
 \institution{Shanghai Key Laboratory of Data Science, School of Computer Science, Fudan University}
 \city{Shanghai}
 \country{China}
}
\email{wdyang@fudan.edu.cn}

\author{Yixuan Li}
\orcid{0000-0002-9229-7555}
\affiliation{%
 \institution{Shanghai Key Laboratory of Data Science, School of Computer Science, Fudan University}
 \city{Shanghai}
 \country{China}
}
\email{yxli24@m.fudan.edu.cn}

\author{Ben Fei}
\orcid{0000-0002-3219-9996}
\affiliation{%
 \institution{Shanghai Key Laboratory of Data Science, School of Computer Science, Fudan University}
 \city{Shanghai}
 \country{China}
 \postcode{200433}
}
\email{bfei21@m.fudan.edu.cn}

\author{Mingjie Zhou}
\orcid{0000-0002-3289-0533}
\affiliation{%
 \institution{Shanghai Key Laboratory of Data Science, School of Computer Science, Fudan University}
 \city{Shanghai}
 \country{China}
 \postcode{200433}
}
\email{mjzhou19@fudan.edu.cn}

\author{Shuhao Li}
\affiliation{%
 \institution{Shanghai Key Laboratory of Data Science, School of Computer Science, Fudan University}
 \city{Shanghai}
 \country{China}
 \postcode{200433}
}
\email{shli23@m.fudan.edu.cn}

\author{Sihang Jiang}
\orcid{0000-0002-0736-6457}
\authornotemark[1]
\affiliation{%
 \institution{Shanghai Key Laboratory of Data Science, School of Computer Science, Fudan University}
 \city{Shanghai}
 \country{China}
 \postcode{200433}
}
\email{jiangsihang@fudan.edu.cn}

\author{Bo Xu}
\orcid{0000-0002-2083-4307}
\affiliation{%
 \institution{School of Computer Science and Technology,
Donghua University}
 \city{Shanghai}
 \country{China}
 \postcode{200433}
}
\email{xubo@dhu.edu.cn}

\author{Yanghua Xiao}
\orcid{0000-0001-8403-9591}
\affiliation{%
 \institution{Shanghai Key Laboratory of Data Science, School of Computer Science, Fudan University}
 \city{Shanghai}
 \country{China}
 \postcode{200433}
}
\email{shawyh@fudan.edu.cn}

\renewcommand{\shortauthors}{L. Ma et al.}

\begin{abstract}
Automated log analysis is crucial to ensure high availability and reliability of complex systems. 
The advent of large language models (LLMs) in natural language processing (NLP) has ushered in a new era of language model-driven automated log analysis, garnering significant interest.
Within this field, two primary paradigms based on language models for log analysis have become prominent. Small Language Models (SLMs) (such as BERT) follow the \textit{pre-train and fine-tune} paradigm, focusing on the specific log analysis task through fine-tuning on supervised datasets. On the other hand, LLMs (such as ChatGPT) following the \textit{in-context learning} paradigm, analyze logs by providing a few examples in prompt contexts without updating parameters. 
Despite their respective strengths, both models exhibit inherent limitations. 
By comparing SLMs and LLMs, we notice that SLMs are more cost-effective but less powerful, whereas LLMs with large parameters are highly powerful but expensive and inefficient.  
To trade-off between the performance and inference costs of both models in automated log analysis, this paper introduces an adaptive log analysis framework known as AdaptiveLog, which effectively reduces the costs associated with LLM while ensuring superior results. This framework collaborates an LLM and a small language model, strategically allocating the LLM to tackle complex logs while delegating simpler logs to the SLM.
Specifically, to efficiently query the LLM, we propose an adaptive selection strategy based on the uncertainty estimation of the SLM, where the LLM is invoked only when the SLM is uncertain.
In addition,  to enhance the reasoning ability of the LLM in log analysis tasks, we propose a novel prompt strategy by retrieving similar error-prone cases as the reference, enabling the model to leverage past error experiences and learn solutions from these cases.
We evaluate AdaptiveLog on different log analysis tasks, extensive experiments demonstrate that AdaptiveLog achieves state-of-the-art results across different tasks, elevating the overall accuracy of log analysis while maintaining cost efficiency.
Our source code and detailed experimental data are available at \url{https://github.com/LeaperOvO/AdaptiveLog-review}.
\end{abstract}




\maketitle

\section{Introduction}

Logs play a pivotal role in maintaining network and software systems by recording crucial runtime information. They serve as essential references for engineers, ensuring the high availability and reliability of complex systems, especially for 24*7 online services like Google, Bing, and Facebook  \cite{le2022log, yu2024deep}. 
However, with the rapid growth volume of logs, it is more challenging for engineers to extract valuable information from massive logs \cite{zhang2021onion}. Consequently, the concept of \textit{automated log analysis} has been proposed, aiming to leverage machine learning (ML) or deep learning (DL) techniques for analyzing logs automatically \cite{he2021survey,le2022log,ma2024influence}. Typical log analysis scenarios include anomaly detection \cite{du2017deeplog,lu2018detecting,zhang2019robust,le2021log}, root cause analysis \cite{shi2023serverrca,wittkopp2024logrca,roy2024exploring}, and failure prediction \cite{zhang2018prefix,gao2020task,cotroneo2019bad}. 
Recently, with the success of pre-trained language models (PLMs) in natural language processing (NLP), especially the advent of large language models (LLMs) represented by ChatGPT and GPT-4 \cite{achiam2023gpt}, LM-based log analysis has gained substantial attention, showcasing remarkable achievements attributed to their powerful understanding and reasoning capability \cite{tao2023biglog,ma2024knowlog,lee2023lanobert,le2023log,xu2024unilog,xu2024divlog,liu2024interpretable,jiang2023lilac}.

Currently, there exist two mainstream paradigms based on language models \footnote{In this paper, Large Language Models (LLMs) refer to models that exhibit universal application and advanced performance with extensive parameter scales without the need for parameter updates. Conversely, Small Language Models (SLMs) are models tailored for specific tasks, characterized by fewer parameters and requiring updates.} in the realm of automated log analysis. The first one relies on small language models (SLMs) represented by BERT \cite{devlin2018bert}, which follow the \textit{pre-train and fine-tune} paradigm, fine-tuning a PLM with annotated task-specific data and thereby enhancing its proficiency in the specific log analysis task \cite{le2023log,chen2022bert,almodovar2024logfit,le2024prelog}.
The fine-tuned SLM can analyze the specific task adeptly and efficiently \cite{ma2024llmparser}.
The second one relies on LLMs represented by ChatGPT, which follow the \textit{in-context learning} (ICL) paradigm, enabling LLMs to learn from a few examples with prompts without updating their parameters. 
By designing high-quality prompts with clear instructions and examples, LLMs with strong reasoning ability analyze logs effectively \cite{xu2024unilog,xu2024divlog}. 

\begin{figure}[]
\centering
   \includegraphics[width=0.75\linewidth]{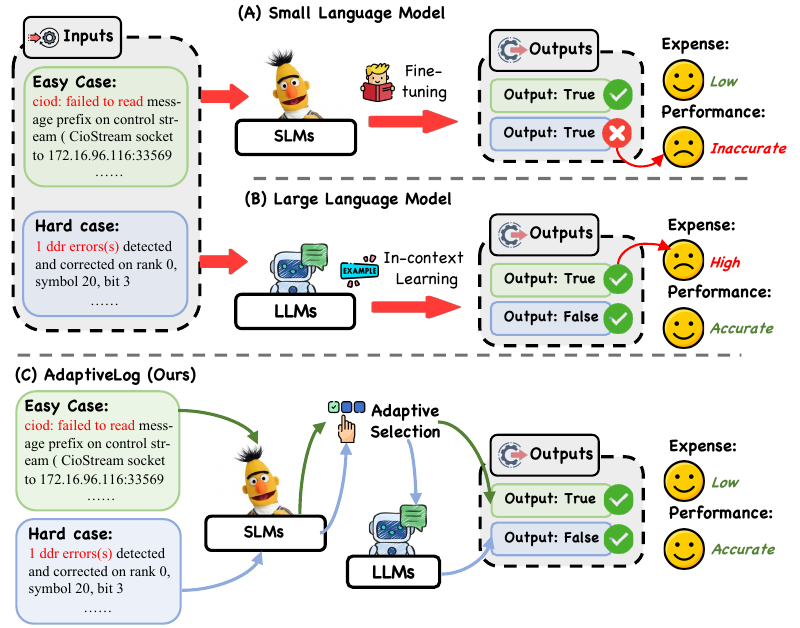}
   \caption{Different log analysis frameworks for anomaly detection with language models, where SLMs are efficient but low performance, and  LLMs are inefficient but high performance. Our proposed framework combines the advantages of both adequately.
}
   \label{fig:overview}
\end{figure}

However, these two prevailing language models for log analysis still exhibit distinct shortcomings. As shown in Fig. \ref{fig:overview}, on the one hand, the capability of the specialized SLM is limited. Considering the diverse and complex datasets encountered in log analysis, 
these SLMs might exhibit sub-optimal performance due to their training limitations, which confine them to specific samples. 
For instance, numerous studies \cite{le2022log,ma2024influence} on anomaly detection indicate that many anomaly patterns are non-typical and occur with low frequency, which hinders the model performance.
On the other hand, LLM-based methods are computationally inefficient and costly due to their complex autoregressive architectures and large number of parameters \cite{frantar2022optq}. For example, modern cloud systems can generate approximately 200 million lines of logs per hour \cite{mi2013toward}. If \textit{GPT-4}\footnote{\url{https://openai.com/api/pricing/}} were used to analyze these logs, with each log message containing at least 10 tokens, the input logs alone would incur an hourly cost of \$60,000, not even accounting for the output costs of \textit{GPT-4}.
By comparing SLMs and LLMs, it becomes apparent that SLMs, with their limited parameters, are more cost-efficient but less powerful. In contrast, LLMs, with their extensive parameters and knowledge, are highly powerful but come with significant costs, including slower inference speed and more expensive prices (APIs) \cite{jiang2023lilac,fu2023specializing,chen2024large}. Hence, can we harness the strengths of both models to trade-off between performance and inference costs in automated log analysis?


To achieve the best of both worlds, this inspired us to exploit these two models working in collaboration, whereby the SLM handles ``simpler samples'', while the LLM takes charge of ``complex samples'' that exceed the former's capabilities. However, there are two key challenges in this process. 
(1) \textbf{Self-Knowledge of the SLM:}  The decision to integrate the LLM stems from recognizing the SLM's limitations in analyzing input logs. 
Insufficient evaluation of the SLM's capabilities can yield substantial influences, including resource misallocation and inefficiency within the LLM. In practice, different log analysis tasks and log samples exhibit varying levels of complexity. For example, the abnormal pattern of logs is diverse in different systems \cite{zhao2021empirical}, making it challenging to define unified simple anomalies that are easier to detect across different systems.
Therefore, ascertaining that the provided logs surpass the capabilities of the SLM poses a challenge.
(2) \textbf{Complex logs Reasoning of the LLM:} Considering that the LLM deals with complex logs that exceed the capability of the SLM,  such logs usually have complex intricate patterns that are hard to capture.
Hence, enhancing the reasoning capabilities of LLMs on complex logs poses another challenge.

To overcome the challenges mentioned above, we propose AdaptiveLog, a novel log analysis framework that focus on the collaboration of LLM and SLM. 
Specifically, AdaptiveLog first predicts the results with the SLM and then queries the LLMs adaptively only if the SLM is uncertain about its results. 
Otherwise, it directly outputs the prediction of the SLM without invoking the LLM.
Rather than analyzing logs only with LLM or SLM, by synergizing the capabilities of both models, AdaptiveLog effectively reduces the costs associated with LLM while ensuring superior results.

Specifically, to solve the first challenge, we propose an adaptive selection strategy to query LLM based on the uncertainty estimation of the SLM.
Many studies \cite{gal2016dropout,kendall2017uncertainties} have indicated that models often exhibit unstable results when their predictions are incorrect.
Building upon this insight, we estimate the uncertainty probability of the SLM based on Bayesian inference. 
Subsequently, the LLM is invoked only when the SLM is uncertain; otherwise, it remains dormant.
To solve the second challenge, motivated by case-based reasoning \cite{kolodner1992introduction,watson1994case} and critical learning pattern of humans \cite{mercer2008talk,reich2023overcome}, we propose a novel prompt strategy by retrieving error-prone cases to empower the reasoning of LLMs for complex log analysis.
Specifically, we gather the error cases of the SLM and analyze these cases with LLM to build a mistake memory bank. During the inference with LLM, we retrieve relevant error cases to assist the LLM in drawing insights from these cases and avoid similar errors. 


To evaluate the effectiveness of AdaptiveLog, we conduct experiments on the software system and network device logs, including six different log analysis tasks. Extensive experiments demonstrate that AdaptiveLog achieves state-of-the-art results across different tasks and enhances the overall accuracy of log analysis while maintaining cost efficiency, allocating the LLM to handle complex logs while efficiently handling simpler logs with the SLM.
In comparison to analyzing all samples with the LLM, AdaptiveLog reduces the costs of LLMs by 73\% while delivering superior results.
In addition, AdaptiveLog exhibits remarkable advantages in low-resource and transfer learning scenarios.

Our main contributions can be summarized as follows:
\begin{itemize}
    \item  We introduce a pioneering practical log analysis framework named AdaptiveLog, which leverages the collaboration of an SLM and an LLM to strike a balance between the performance and inference costs of language models. Our approach is simple and non-parametric to harness the distinct strengths of both models effectively.
    \item We propose an adaptive selection strategy to query the LLM based on uncertainty estimation, allocating the LLM to handle complex logs, while efficiently handling simpler logs with the SLM.
    Moreover, we propose a novel prompt strategy by retrieving error-prone cases to enhance the reasoning capabilities of the LLM for complex log analysis, contributing to more accurate and insightful log analysis outcomes.
    \item We extensively evaluate AdaptiveLog on six different log analysis tasks. Our evaluation results underscore the framework's high effectiveness and efficiency, demonstrating its ability to strike a balance between performance and inference costs in log analysis scenarios.
\end{itemize}


\section{Related Works}
\subsection{Large Language Models}
Large language models have significantly advanced the field of NLP, leveraging extensive training corpora and computational resources \cite{kasneci2023chatgpt}. Notably, LLMs such as ChatGPT and GPT-4 \cite{achiam2023gpt}, with their vast parameter scales and alignment with human feedback, have opened new avenues for software engineering \cite{wang2024software,li2023exploring,fan2023automated,du2024evaluating,guo2024deepseek}. With the significant increase in parameter scale and corpus size, LLMs have become more and more powerful. Researchers have discovered that these capabilities can be effectively harnessed through textual prompts \cite{brown2020language,sahoo2024systematic}. Hence, ICL has emerged as a prominent research topic, enabling LLMs to adapt to new tasks without parameter updates by conditioning on a few labeled examples \cite{dong2022survey,wei2023larger}. This approach bypasses the need for model training and achieves impressive performance by utilizing a prompt structure that typically includes: (1) an instruction specifying the task, (2) similar examples with ground truth to impart task-specific knowledge, and (3) a query expecting an informed response from the LLM. Many studies have shown that LLMs excel in addressing complex problems, such as code generation \cite{liu2024your,du2024evaluating, yu2024codereval}, automated program repair \cite{jiang2023impact,hossain2024deep,zirak2024improving}, and mathematical reasoning \cite{ahn2024large,zhang2024evaluating}, within the ICL framework. The underlying principle of ICL is believed to be the complex reasoning ability acquired by LLMs through extensive pre-training \cite{minaee2024large}. Hence, this enables them to solve the given problem with human-like logic based on the analogy with demonstration examples in the prompt. 

Although LLMs provides new opportunities to solve complex problems, LLMs still face the issue of high costs and inefficiency in inference due to the huge number of parameters \cite{xu2024survey,gu2024minillm}, making them less attainable for individuals and smaller organizations. In addition, many studies \cite{gao2023makes,ma2023large} have shown that examples are crucial for ICL and directly affect the inference results of LLMs. 
Considering the complexity of contextual information in logs, analyzing logs with ICL becomes more challenging.
In this paper, we focus on automated log analysis based on LLMs and propose a new framework with the collaboration of large and small language models to reduce the cost of LLMs. Further, we propose a novel prompt strategy with error-prone cases to enhance LLM's reasoning ability in log analysis and avoid the same errors.

\subsection{Automated Log Analysis with Language Models}
The success of pre-trained language models (PLMs) and large language models (LLMs) in natural language processing (NLP) has inspired significant advancements in automated log analysis. This research can be broadly categorized into two main approaches. One type of research is based on the pre-trained language model represented by BERT \cite{devlin2018bert}, these works directly represent logs with the pre-trained model or fine-tune on specific datasets to analyze logs. HitAnomaly \cite{huang2020hitanomaly}, SwissLog \cite{li2020swisslog}, NeuralLog \cite{le2021log} and ClusterLog \cite{egersdoerfer2022clusterlog} utilize the pre-trained model to obtain the embedding of logs for log-based anomaly detection. Following the \textit{pre-train and fine-tune} paradigm of the pre-trained model, BertLog \cite{chen2022bert}, LogEncoder \cite{qi2023logencoder}, LogPPT \cite{le2023log}, HilBERT \cite{huang2023improving}, Lanobert \cite{lee2023lanobert}, LogFiT \cite{almodovar2024logfit} and PreLog \cite{le2024prelog} utilize different fine-tuning strategies to train the model on specific log analysis tasks, which significantly improves the ability of log analysis. In addition, these log analysis works are mainly based on general pre-trained language models, and due to the differences between natural language and logs, many works further enhance the pre-trained model's log understanding ability by pre-training on log corpus. Biglog \cite{tao2023biglog} has been proven to handle complex log analysis tasks better by learning in-sentence and cross-sentence features of logs. KnowLog \cite{ma2024knowlog} first proposes leveraging domain knowledge to enhance log pre-training, which enables further improvement of the model's log understanding.

The other type of research is based on large language models represented by ChatGPT and GPT-4, these works draw on the in-context learning (ICL) capabilities of LLMs to analyze logs without updating the parameters.
LLMParser \cite{ma2024llmparser} first evaluates the performance of different LLMs on log parsing and finds that LLMs significantly outperform other state-of-the-art parsers. In addition, LogPrompt \cite{liu2024interpretable}, DivLog \cite{xu2024divlog}, UniLog \cite{xu2024unilog}, SCLogger \cite{li2024go} and LILAC \cite{jiang2023lilac} also demonstrate the effectiveness of log analysis using LLMs.
Despite the superior performance of LLMs they also face the challenge of inefficiency and high cost, and the performance of the smaller pre-trained model is relatively weak but cost-efficient. LLMParser \cite{ma2024llmparser} experimentally verified that the smaller language model is more efficient than complex LLMs. This paper introduces a pioneering log analysis framework designed to fully exploit the advantages of both large and small language models.

\begin{figure*}[]
\centering
   \includegraphics[width=\linewidth]{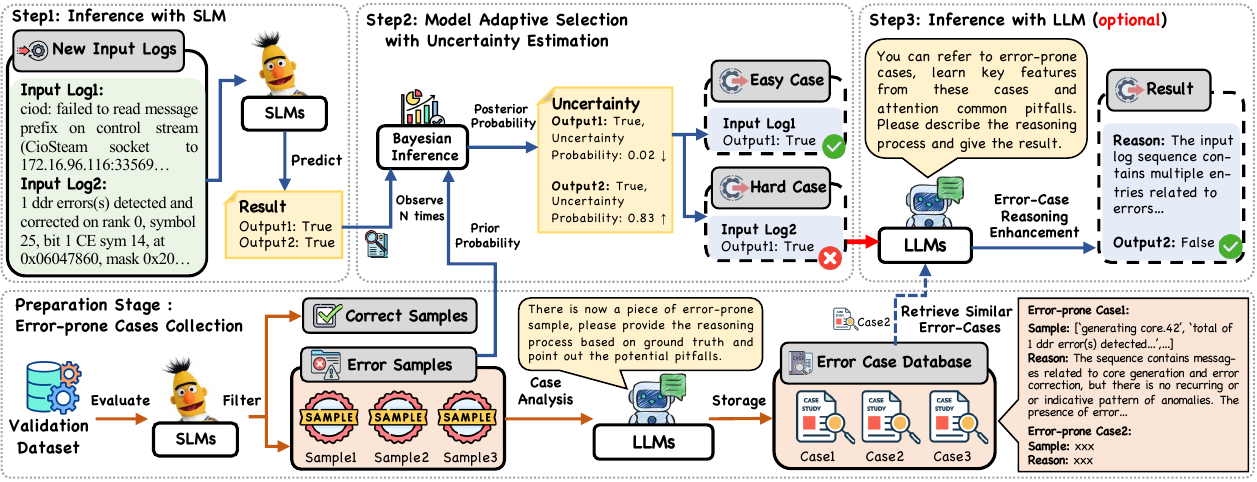}
   \caption{The framework of AdaptiveLog, which comprises three key steps and a preparation stage. It first analyzes logs with the SLM and adaptively chooses to invoke the LLM based on the uncertainty of the SLM. The preparation stage constructs an error-prone case database to enhance the reasoning of the LLM. }
   \label{fig:framework}
\end{figure*}

\section{Methods}
\subsection{Overview}
Fig. \ref{fig:framework} shows the framework of AdaptiveLog, which consists of three key steps when inference.  Specifically, given a specific log analysis task and a corresponding SLM, the SLM first gives its prediction in the first step. Then, the uncertainty of SLM is estimated through Bayesian inference to adaptively select the LLM in the second step, which would give a probability of uncertainty. Finally, when the probability of uncertainty is higher, AdaptiveLog will query the LLM to analyze logs in the third step, otherwise, it will directly output the results of the SLM. To enhance the reasoning ability of the LLM on complex logs, we propose a novel error-case reasoning enhancement prompt strategy by retrieving similar error-prone cases as the reference, which can leverage past error experience and learn solutions from these cases. 
Accordingly, the framework also includes a preparation stage for the collection of error-prone cases, where we collect error samples of the SLM on this task and then utilize the LLM to analyze their reasoning processes about ground truth and potential pitfalls susceptible to errors.
Following this, this information is structured into the error-prone cases database within a key-value datastore to support the reasoning of the LLM.
In the following sections, we first describe the preparation stage of AdaptiveLog, and then describe the details of uncertainty estimation and the prompt strategy for reasoning. 


\subsection{Error-prone Cases Collection}
Given a SLM $\mathcal{M}_{SLM}$ for a specific log analysis task, to analyze why error samples tend towards inaccuracies and to avoid errors again when reasoning with the LLM $\mathcal{M}_{LLM}$, we construct an error-prone case database $\mathcal{D}$, which serves as a foundation to bolster the reasoning of the LLM.  This database is created by filtering the logs where the SLM goes wrong and analyzing them based on the LLM, finally storing them as error-prone cases, consisting of the raw log and the detailed analysis.

Firstly, given a task-specific validation set $\mathcal{V} = \{(x_i, y_i)\}$ and the SLM $\mathcal{M}_{SLM}$, $\mathcal{M}_{SLM}$ is then filtered for 
correct samples $\mathcal{S}_{Correct} = \{(x_i,y_i) | (x_i, y_i) \in V, \mathcal{M}_{SLM}(x_i) = y_i  \}$  
and error samples $\mathcal{S}_{Error}= \{(x_i,y_i) | (x_i, y_i) \in V, \mathcal{M}_{SLM}(x_i) \neq y_i  \}$ on $\mathcal{V}$, where $x_i$ is the input log and $y_i$ is the ground-truth. Here, we can calculate the error rate $err$ of $\mathcal{M}_{SLM}$, ||·|| represents the set size:
\begin{equation}
    \mathit{err} = \frac{||\mathcal{S}_{Error}||}{||\mathcal{V}||}. \label{equation1}
\end{equation}

Secondly, we employ the LLM $\mathcal{M}_{LLM}$ to analyze the reasoning process about ground truth and potential pitfalls of the error samples $\mathcal{S}_{Error}$. Specifically, given an error sample $(x_i, y_i) \in \mathcal{S}_{Error}$ and the instruction $\mathcal{I}_{Reason}$,  $\mathcal{M}_{LLM}$ gives the reasoning process based on the sample's inputs $x_i$ and ground-truth $y_i$. The content of this instruction as shown in Fig. \ref{fig:framework} (1) is designed to make the LLM analyze as desired. Finally, we store the original sample $x_i$ as key, and the analysis result $\mathcal{R}$ of the LLM as value in the database $\mathcal{D}$ as a key-value store. Hence, the error-prone case database is constructed and will support the subsequent reasoning of the LLM. We provide an error case example in Fig. \ref{fig:prompt} to understand the form of the constructed error case intuitively. 

\begin{gather}
    \mathcal{R}_i = \mathcal{M}_{LLM}(\mathcal{I}_{Reason},x_i,y_i) \notag, \\
    \mathcal{D} = \{(x_i, R_i) | (x_i, y_i) \in \mathcal{S}_{Error} \}.
\end{gather}

\subsection{Uncertainty Estimation}
When given new input $x$, the SLM first gives the prediction $y' = arg max(p_{SLM}(y | x))$, where $y'$ represents the predicted category with the highest
probability. Then we estimate the uncertainty probability of model $p(C | (x, y'))$  conditional on the given input $x$ and the prediction $y'$, where event $C$ indicates that the model is uncertain about the result. However, $p(C | (x, y'))$ cannot be calculated directly, we employ Bayes' rule to solve, that is:
\begin{equation}
    p(C | (x, y')) \propto p(C) \times p((x, y') | C), 
\end{equation}
where $p(C)$ represents the prior probability. Importantly, the result of the model is usually incorrect when the model is uncertain \cite{gal2016dropout,kendall2017uncertainties}. Assuming that observing different inputs are independent and identically distributed (i.i.d.), we can use the error rate $err$ (Equation \ref{equation1}) as the prior probability $p(C) = err$. The likelihood probability $p((x, y') | C)$ indicates the probability of observing $(x, y')$ given the event $C$. We can observe the predictions $y'$ of the model multiple times by giving the same $x$. If there is a difference in the predictions, it is classified as event $C$. Bayesian inference utilizes these new observations to update the prior probability. 

We perform $N$ times predictions for the same input as the newly considered observations. Assuming the event $C$ occurs $\alpha$ times, then $N - \alpha$ times the model is certain. The likelihood probability $p((x, y') | C)$ can be calculated as follows:
\begin{equation}
    p((x, y') | C) \approx p(C)^\alpha(1 - p(C))^{N-\alpha}.
\end{equation}

Based on the Bayesian inference, $p(C)$ is a distribution represented by $\theta = p(C)$. Considering that the expectation of this distribution is equal to $err$, it is reasonable to consider $p(\theta) \sim Beta(err, 1 - err)$. The selection of the Beta distribution on the one hand because of its flexibility and suitability for modeling probabilities between 0 and 1, and on the other hand it is convenient to compute since it belongs to the conjugate prior distribution \cite{joyce2003bayes}. Hence, based on the new observation on $(x, y')$, we can deduce the posterior distribution $p(\theta | (x, y'))$: 
\begin{equation}
p(\theta | (x, y')) \sim Beta(err + \alpha, (1 - err) + N - \alpha).  
\end{equation}

At last, the calculation of $p((x, y') | C)$ can be formulated as:
\begin{align}
    p(C | (x, y')) & = E(p(\theta | (x, y'))) \notag \\
    & = \frac{err + \alpha}{err + \alpha + (1 - err) + N - \alpha} \notag \\
    & = \frac{err}{N + 1} + \frac{\alpha}{N + 1},
    \label{bayes}
\end{align}
where the first term is a constant and the second term is determined from observations. Following Monte Carlo Dropout \cite{gal2016dropout}, we determine whether the uncertain event $C$ has occurred by comparing it with the model's random range on correct samples. Specifically, we first calculate the range of probability variations of the correct samples $S_{Correct} = \{(x_i,y_i)\}_{i=1}^M$ on the validation set:
\begin{gather}
    Mean =  \frac{1}{N} \sum_{n=1}^N p(y_i|x_i,\delta_n) \notag, \\
    Variation =  \frac{1}{M} \sum_{i=1}^M \frac{1}{N} \sum_{n=1}^N \mid p(y_i|x_i,\delta_n) - Mean \mid,
\end{gather}
where we perform $N$ passes of forward propagation through the SLM on the same sample, $N$ is empirically set to 10 in our experiments. In each pass, part of neurons in model $\delta$ are randomly deactivated, $p(y_i|x_i,\delta_n)$  denotes the probability of the result at the $n$ pass. 

Then we perform the same $N$ passes on the new input to be estimated, and quantify the variation of probability each time:
\begin{equation}
    Obseration_n = \mid p(y'|x,\delta_n) - \frac{1}{N} \sum_{n=1}^N p(y'|x,\delta_n)  \mid.
\end{equation}

Finally, we determine the $\alpha$ by counting the number of $Obseration_n <= Variation$ in $N$ observations. For example, suppose the error rate $err$ of the SLM is 0.2, and given a new sample $x$, 5 out of 10 observations are uncertain. Based on Equation \ref{bayes}, the probability of uncertainty is 0.472.

After calculating the probability of model uncertainty $p(C | (x, y'))$, then the probability of model certainty equals $1 - p(C | (x, y'))$. We specify that the LLM is queried only if the probability of model uncertainty is higher than that of certainty, where $\mathbbm{I(\cdot)}$ denotes the indicator function:
\begin{equation}
    Select\_LLM = \mathbbm{I} [ p(C | (x, y')) > 1 - p(C | (x, y')) ].\label{select}
\end{equation}

\subsection{Error-Case Reasoning Enhancement Strategy}
Based on Equation \ref{select}, AdaptiveLog performs adaptive model selection.
The LLM is invoked when the probability of model uncertainty surpasses that of model certainty; otherwise, it directly yields the output of the SLM without querying the LLM.
We designate samples with low uncertainty that do not necessitate the invocation of the LLM as \textbf{simple samples}, otherwise as \textbf{hard samples}.
Since hard samples are error samples of the SLM, understanding these error-prone logs requires extensive experience and knowledge compared to simple logs.
To improve the reasoning of the LLM over complex logs, inspired by the classical AI paradigm case-based reasoning \cite{kolodner1992introduction,watson1994case} and critical
learning pattern of humans \cite{mercer2008talk,reich2023overcome}, which can refer to useful experiences and learn from mistakes to avoid making them again, we propose a novel prompt strategy called \underline{E}rror-\underline{C}ase \underline{R}easoning Enhancement (ECR) employing error-prone cases to enhance the reasoning of the LLM. ECR solves the complex log analysis task by retrieving similar error-prone cases, referring to their reasoning process and noticing the potential pitfalls susceptible to errors.

As shown in Fig. \ref{fig:framework} (3), ECR based on LLMs first retrieves similar cases from the error-prone case database $\mathcal{D}$ to construct the prompt, and then the LLM follows the prompt's instructions to reason. By referring to these cases, the LLM first gives the reasoning process and then outputs the result. Now we elaborate on the error-prone case selection and prompt strategy as follows.

\subsubsection{Error Case Selection}
Since LLMs are not inherently trained for log analysis, they lack precise knowledge to analyze these complex logs. To enable the LLM to learn how to analyze logs and avoid the same mistakes, ECR retrieves relevant cases from database $\mathcal{D}$ that are similar to the current input logs. Specifically, we first embed the input log $x$ and case $c_i \in \mathcal{D}$ from the raw data into the vector representation $\mathbf{E}(x), \mathbf{E}(c_i)$. For case $c_i$ from the key-value store database $\mathcal{D}$, we only calculate the embedding of the key. Then, we calculate the cosine similarity $sim(x,c_i)$ between x and all candidate cases with the following equation, where $\mathbf{E(\cdot)}$ denotes the pre-trained embedding model. Finally, the top-$k$ cases with the highest similarities are retrieved. $k$ is set to 5 in our works.
\begin{equation}
    sim(x,c_i) = \cos (\mathbf{E}(x), \mathbf{E}(c_i)) = \frac{\mathbf{E}(x) \cdot \mathbf{E}(c_i)}{||\mathbf{E}(x)|| ||\mathbf{E}(c_i)||}.
\end{equation}

\begin{figure}[]
\centering
   \includegraphics[width=0.88\linewidth]{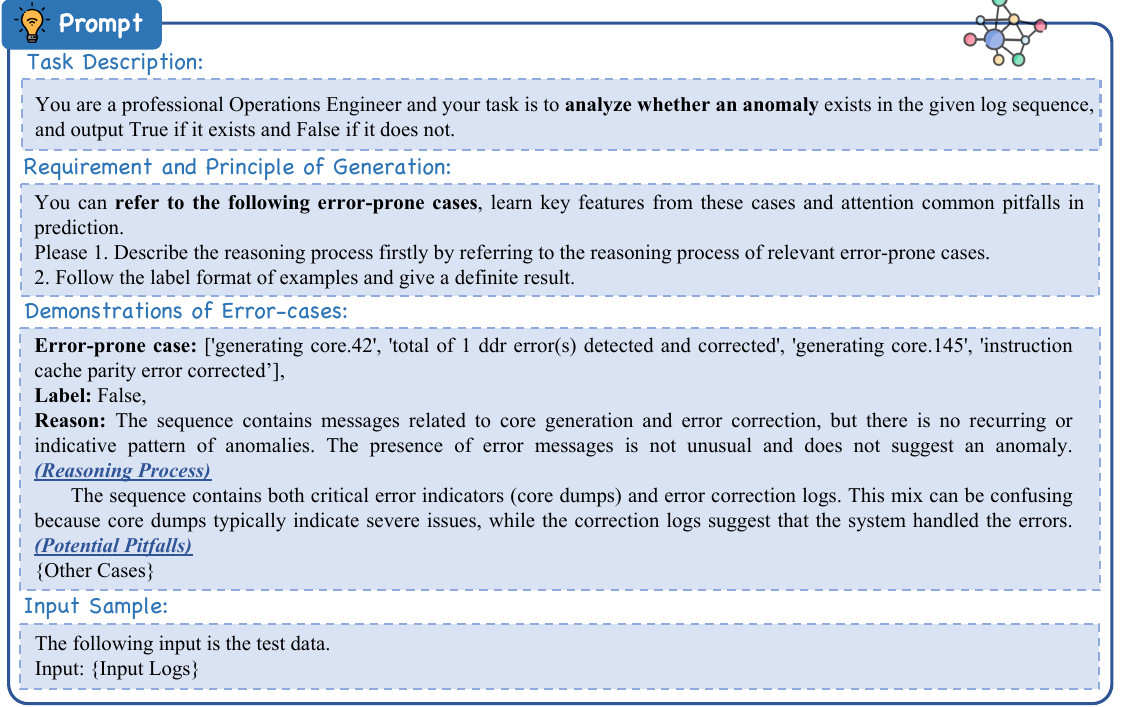}
   \caption{The prompt template example of ECR on the anomaly detection task, where the error-prone cases including reasoning process and potential pitfalls are provided in the prompt to improve the reasoning of LLMs.}
   \label{fig:prompt}
\end{figure}

\subsubsection{Prompt Strategy}
Designing an appropriate prompt is the most crucial part of the error-case reasoning enhancement strategy. As shown in Fig. \ref{fig:prompt}, a carefully designed prompt for ECR including task description, requirement, selected similar cases, and input logs.

First, the task description is designed to introduce the specific log analysis task to be performed. 
Next, we require the LLM to refer to the reasoning process of these error-prone cases and notice the potential pitfalls. By referring to these cases, the LLM effectively learns solutions and insights from them to analyze logs. This allows the LLM to align with the capabilities of domain experts. The LLM is required to give the reasoning process first, and then output the result. This facilitates not only the learning of the reasoning process but also the explicit understanding of the LLM's reasoning process.
In addition, to reduce the redundant contents in the raw output, we require the LLM to follow the format of similar cases. Then we can easily utilize regular expressions to extract the result in the output.
For selected similar cases, many studies \cite{kumar2021reordering, zhao2021calibrate,xu2024divlog} have demonstrated that the permutation of various examples in the context can significantly impact performance. For example, \textit{Zhao et al.} \cite{zhao2021calibrate} highlighted that the model's prediction for a query may exhibit bias toward the most recent example (referred to as recency bias). This implies that if the closest example to the query within the prompt bears sufficient similarity to the query itself, the model's predictions will incline toward outcomes proximal to the query.
Hence, following these findings, we arrange these examples in \textit{ascending order} based on the cosine similarity. 
To visually understand ECR's prompt template, Fig. \ref{fig:prompt} shows a complete prompt example of the anomaly detection task.

\section{Experiments}

\subsection{Environment and Implementation}
In our experiments, we choose the widely used BERT \cite{devlin2018bert} as the SLM. Although the pre-trained model for logs, such as Biglog \cite{tao2023biglog} yields better performance than BERT, we choose the general pre-trained model BERT given the extra expense of pre-training on logs and BERT is more available in practice.
Given the specific log analysis task,  we first fine-tune BERT on the training set to adapt it to the specific task. During fine-tuning, we set the batch size as 16, and epochs as 3. Moreover, the optimizer we adopt is Adam with a learning rate of 5e-5 and the linear schedule with a warmup ratio of 0.01.

Considering the performance and price of the LLM, we choose the most popular ChatGPT (\textit{gpt-3.5-turbo-16k}) due to its popularity in the community to analyze error cases and perform ECR. We utilize  HTTP requests to invoke the OpenAI APIs and interact with ChatGPT. 
In addition, the embedding model used for sample selection in ICL and ECR we choose \textit{bge-large-en-v1.5}, due to its excellent performance among all open-source embedding models. The demonstration examples of ICL are selected from the training set and set to 5.
To increase the stability of LLM’s output and ensure reproducibility for the same query, we set the \textit{temperature} of ChatGPT to 0.
We conduct all the experiments on 4 NVIDIA RTX 3090 GPUs.

\subsection{Downstream Tasks and Datasets}

To verify the usability and generalizability of AdaptiveLog on different log analysis tasks, we conduct experiments on six different downstream tasks across different domains, including software system and network device logs. 
Following existing studies in software system logs \cite{he2021survey,zhu2023Loghub}, we conduct experiments on Anomaly Detection and Failure Identification, two of the most widely studied tasks in the log analysis field. 
Considering that existing public datasets for these tasks often exhibit relatively simple anomaly patterns \cite{zhao2021empirical,yu2024deep}, we also conduct experiments on network device logs to verify the generalizability across different log analysis domains. 
These tasks require domain knowledge and can effectively evaluate model capabilities, a prerequisite for tackling other downstream tasks, posing significant challenges.
Following the task construction process on network device logs \cite{ma2024knowlog}, we construct four different log analysis tasks across two vendors, Cisco\footnote{\url{https://www.cisco.com/c/en/us/support/all- products.html}} and Huawei\footnote{\url{https://support.huawei.com/enterprise/en/index.html}}, including two devices: Switches and Routers. 
We provide statistics for different tasks of their datasets in Table \ref{tab: dataset_software} and \ref{tab:dataset_device}. Next, we give an introduction to each task and its evaluation metrics.

\begin{table}[]
\centering
\caption{Statistics of downstream tasks datasets on software systems \\(Training / Validation / Testing size).}
\label{tab: dataset_software}
\resizebox{0.5\columnwidth}{!}{%
\begin{tabular}{l|lc}
\toprule[1.5pt]
Tasks                              & Dataset     & \#Size            \\ \midrule
\multirow{2}{*}{Anomaly Detection} & BGL         & 6,000 / 2,000 / 2,000 \\
                                   & ThunderBird & 6,000 / 2,000 / 2,000 \\ \midrule
Failure Identification             & Openstack   & 196 / 100 / 100       \\ \bottomrule[1.5pt]
\end{tabular}%
}
\end{table}

\begin{table}[]
\centering
\caption{Statistics of downstream tasks datasets on network devices \\(Training / Validation / Testing size).}
\label{tab:dataset_device}
\resizebox{0.68\columnwidth}{!}{%
\begin{tabular}{l|lcc}
\toprule[1.5pt]
Tasks                                                                                            & Vendors/Devices & Switches          & Routers           \\ \midrule
\multirow{2}{*}{\begin{tabular}[c]{@{}l@{}}Module\\ Classification\end{tabular}}                 & Cisco           & 536 / 179 / 178       & 1,244 / 415 / 414     \\
                                                                                                 & Huawei          & 594 / 198 / 197       & 427 / 142 / 142       \\ \midrule
\multirow{2}{*}{\begin{tabular}[c]{@{}l@{}}Level \\ Prediction\end{tabular}}                     & Cisco           & 2,339 / 780 / 779     & 2,369 / 789 / 789     \\
                                                                                                 & Huawei          & 1,833 / 611 / 610     & 1,257 / 419 / 418     \\ \midrule
\multirow{2}{*}{\begin{tabular}[c]{@{}l@{}}Log and Description\\ Semantic Matching\end{tabular}} & Cisco           & 7,199 / 2,400 / 2,399 & 6,309 / 2,103 / 2,102 \\
                                                                                                 & Huawei          & 7,342 / 2,447 / 2,447 & 5,477 / 1,826 / 1,825 \\ \midrule
\begin{tabular}[c]{@{}l@{}}Log and Possible \\ Cause Ranking\end{tabular}                        & Huawei          & 2,904 / 984 / 988     & 2,311 / 778 / 720     \\ \bottomrule[1.5pt]
\end{tabular}%
}
\end{table}

\subsubsection{Downstream Tasks of Software System Logs}
\begin{enumerate}
    \item \textbf{Anomaly Detection (AD).} AD is a widely researched log analysis task to predict whether anomalies exist within a short period of log messages. We take the log sequence as input, then the model analyzes the anomaly in a sequence.
    
    \underline{\textit{Dataset and Metric.}} We evaluate the anomaly detection performance on two datasets (BGL and ThunderBird) from Loghub \cite{zhu2023Loghub}, which contributes large-scale system log datasets to the community. BGL and ThunderBird are selected based on the simplicity of anomaly
patterns in other public Anomaly Detection datasets \cite{zhao2021empirical,yu2024deep}.  For example, BERT achieves 100.0 / 100.0 / 100.0 Precision / Recall / F1-value on HDFS, which is insufficient for evaluating AdaptiveLog.
Given the extensive size of BGL and Thunderbird, we take 200,000 raw logs in chronological order with a window size of 20, following previous research 
\cite{le2022log}, these logs are grouped into datasets (6,000 / 2,000 / 2,000) chronological selection to prevent data leakage. 
Following Biglog \cite{tao2023biglog}, we concatenate each log in the sequence and then input the language model.
    Following previous anomaly detection works \cite{le2021log,le2022log,tao2023biglog}, we use Precision, Recall, and F1 on the anomaly class as evaluation metrics.

    \item \textbf{Failure Identification (FI).} Unlike anomaly detection identifying present faults from logs, FI further identifies what type of failure occurs in the anomaly log.
    
    \underline{\textit{Dataset and Metric.}} We collect this dataset from \cite{cotroneo2019bad}, which is an OpenStack dataset including 396 failure tests and 16 kinds of API errors, such as ``server add volume error''. Different from the binary classification task of anomaly detection, this task is a 16-class classification task for 16 error types in the dataset. Considering that engineers are interested in whether the top-K recommended results contain the correct error, we report the Recall@k rate as the evaluation metric.
\end{enumerate}

\subsubsection{Downstream Tasks of Network Device Logs}

\begin{enumerate}
    \item \textbf{Module Classification (MC).} MC is a multi-class classification task aiming at identifying which module the log originates from, where the model needs to understand the contextual information of the logs to accurately identify their source.
    
    \underline{\textit{Dataset and Metric.}} We collect network device logs from public documentation and replace the module name of raw logs with \textit{[MASK]} as input, the module name in the log as ground truth. 
    As the unbalanced multi-class classification task and considering the importance of different classes, we report Accuracy and Weighted F1 as evaluation metrics.
    
    \item \textbf{ Level Prediction (LP).} LP is a binary classification task designed to determine the risk level of the given log: \textit{ERROR} or \textit{INFO}. The \textit{ERROR} level indicates that the log has a serious impact on the system, while the \textit{INFO} level indicates that the log is informational only.
    
   \underline{\textit{Dataset and Metric.}}
    We construct datasets based on the collected network device logs, referring to \cite{xu2024unilog}, we replace the verbosity level of the original logs with \textit{[MASK]} as input, where the original verbosity level is ground truth. 
    Similar to the anomaly detection task, we take Precision, Recall, and F1 on the ERROR class as evaluation metrics.
    
    \item \textbf{Log and Description Semantic Matching (LDSM).}
    LDSM is a matching task to determine whether the given log is semantically consistent with the natural language description.
    This task aims to evaluate the model's ability to understand the semantics of logs.
    
    \underline{\textit{Dataset and Metric.}} We collect logs and their corresponding descriptions from public documentation, establishing pairs of [Log, Description] as the ground truth. Subsequently, we randomly choose one other description for each log as a negative sample. As a binary classification task, both positive and negative cases require attention, we present Accuracy and Weighted-F1 as evaluation metrics.
    
    \item \textbf{Log and Possible Cause Ranking (LPCR).} LPCR is a ranking task to find the possible cause of the given log from multiple candidates. This log analysis task is designed to evaluate whether the model has strong professional background knowledge.
    
    \underline{\textit{Dataset and Metric.}} We collect logs and their possible causes from the Huawei public documentation to establish pairs of [Log, Possible Cause] as the ground truth. Subsequently, we randomly pick four causes of other logs to form a candidate set. As a typical ranking task, following existing works \cite{niu2022spt,shi2023cocosoda}, we report Precision@K and Mean Reciprocal Rank (MRR) as evaluation metrics, where MRR is a statistic measure for evaluating search algorithms.
    
\end{enumerate}

\subsection{Baselines}
We categorize baselines for log analysis into three groups according to technology type: traditional deep-learning methods (\textbf{CNN} \cite{lu2018detecting}, \textbf{BiLSTM} \cite{zhang2019robust}), pre-trained language models (\textbf{BERT} \cite{le2021log,lee2023lanobert}, \textbf{Biglog} \cite{tao2023biglog}, \textbf{KnowLog} \cite{ma2024knowlog}), and large language models (\textbf{ChatGPT}). 
\textbf{CNN} and \textbf{BiLSTM} convert each log message into a vector with the word embedding model and then input the vector to a deep neural network (CNN or BiLSTM) to analyze logs. 
As a general pre-trained language model, \textbf{BERT} is the first and most widely used pre-trained model for log analysis. It can represent logs well due to its excellent semantic representation.
As a pre-trained model specifically designed for logs, \textbf{Biglog} excels in capturing essential features due to its extensive training on the log corpus. 
\textbf{KnowLog} is the state-of-the-art log pre-trained model in log analysis due to pre-training on log corpus with knowledge enhancement, equipping it with the domain knowledge needed to understand logs.
For fairness, we reproduce Biglog and KnowLog with the same pre-training setting on our log corpus (all training sets). Following \textit{pre-train and fine-tune} paradigm, all pre-trained models are fine-tuned in log analysis tasks with corresponding training sets.
\textbf{ChatGPT} has demonstrated amazing application potential in multiple domains with its excellent natural conversation capabilities and extensive knowledge base. Following \textit{in-context learning} paradigm, we select top-5 examples from the training set for each input log and use the same embedding model as ECR to select examples.

In addition, to validate the effectiveness of our proposed ECR strategy, we also compare it with the ICL ability of LLMs to analyze the hard samples. Hence, we also report the results of analyzing hard samples with ICL in our experiments.

\subsection{Evaluation}
We evaluate AdaptiveLog by answering the following research questions (RQs):

\begin{table}[]
\centering
\caption{Results on Anomaly Detection and Failure Identification. Note: \textit{Easy (Hard) samples} represent low (high)-uncertainty samples.}
\label{tab:main_exp1}
\resizebox{0.65\columnwidth}{!}{%
\begin{tabular}{lcc|c}
\toprule[1.5pt]
Models            & \multicolumn{2}{c|}{AD (Precision / Recall / F1)}                                                                                          & FI (Recall @ 1 / 2)                                               \\ \midrule 
\begin{tabular}[c]{@{}l@{}}Percentage of \\ hard samples\end{tabular}                 & \begin{tabular}[c]{@{}c@{}}BGL\\ 27.7\%\end{tabular} & \begin{tabular}[c]{@{}c@{}}ThunderBird\\ 29.4\%\end{tabular} & \begin{tabular}[c]{@{}c@{}}OpenStack\\ 38\%\end{tabular} \\ \midrule
CNN               & 99.89 / 93.84 / 96.77                                            & 81.95 / \textbf{100.0} / 90.08                                                   & 79.0 / 87.0                                                       \\
BiLSTM            & 99.35 / 96.97 / 98.15                                           & \textbf{96.98} / 89.12 / 92.89                                                   & 80.0 / 91.0                                                       \\
Biglog            & \textbf{100.0 } / 97.49 / 98.73                                            & 95.76 / 90.65 / 93.13                                                   & 82.0 / 94.0                                                       \\
KnowLog           & 100.0 / 97.49 / 98.73                                                               & 96.70 / 89.33 / 92.87                                                                       & 82.0 / 95.0                                                                 \\ \midrule
BERT              & \textbf{100.0 } / 97.49 / 98.73                                           & 95.27 / 90.53 / 92.84                                                   & 83.0 / 92.0                                                       \\
\quad $\mapsto Easy \: Samples$      & \textit{100.0 / 99.68 / 99.84}                                           & \textit{99.16 / 96.57 / 97.85}                                                & \textit{88.7 / 92.0}                                                       \\
\quad  $\mapsto Hard \: Samples$      & \textit{0.0 / 0.0 / 0.0 }                                                & \textit{69.78 / 57.21 / 62.87 }                                                  & \textit{73.7 / 92.0 }                                                    \\ \midrule
ChatGPT           & 96.04 / 98.85 / 97.42                                                               & 83.93 / 100.0 / 91.26                                                                       & 86.0 / \textbf{96.0}                                                       \\ \midrule
AdaptiveLog (ICL) & 98.14 / 99.48 / 98.81                                           & 86.14 / 97.09 / 91.29                                                   & 86.0 / 93.0                                                       \\
\quad  $\mapsto Hard \: Samples$      & \textit{51.35 / 90.48 / 65.52}                                           & \textit{50.68 / 100.0 / 67.27}                                                   & \textit{81.6 / 94.7}                                                       \\ \midrule
AdaptiveLog (ECR) & 98.15 / \textbf{99.68} / \textbf{98.91}                                           & 93.04 / 97.03 / \textbf{94.99}                                                   & \textbf{87.0} / 94.0                                                       \\
\quad  $\mapsto Hard \: Samples$      & \textit{53.85 / 100.0 / 70.0}                                            & \textit{69.93 / 99.54 / 82.16}                                                   & \textit{84.2 / 97.3}                                                       \\ \bottomrule[1.5pt]
\end{tabular}%
}
\end{table}

\begin{table}[]
\centering
\caption{Results on Module Classification and Level Prediction.}
\label{tab:main_exp2}
\tabcolsep=0.08cm
\resizebox{\columnwidth}{!}{%
\begin{tabular}{lcccc|cccc}
\toprule[1.5pt]
Models                                                               & \multicolumn{4}{c|}{Module Classification (Accuracy / Weighted-F1)}                                                                                                                                                                                              & \multicolumn{4}{c}{Level Prediction (Precision / Recall / F1)}                                                                                                                                                                                                    \\ \midrule
                                                                     & \multicolumn{2}{c|}{Cisco}                                                                                                                & \multicolumn{2}{c|}{Huawei}                                                                                          & \multicolumn{2}{c|}{Cisco}                                                                                                                & \multicolumn{2}{c}{Huawei}                                                                                           \\
\begin{tabular}[c]{@{}l@{}}Percentage of\\ hard samples\end{tabular} & \begin{tabular}[c]{@{}c@{}}Switches\\ 29.2\%\end{tabular} & \multicolumn{1}{c|}{\begin{tabular}[c]{@{}c@{}}Routers\\ 27.8\%\end{tabular}} & \begin{tabular}[c]{@{}c@{}}Switches\\ 28.2\%\end{tabular} & \begin{tabular}[c]{@{}c@{}}Routers\\ 25.4\%\end{tabular} & \begin{tabular}[c]{@{}c@{}}Switches\\ 17.5\%\end{tabular} & \multicolumn{1}{c|}{\begin{tabular}[c]{@{}c@{}}Routers\\ 16.1\%\end{tabular}} & \begin{tabular}[c]{@{}c@{}}Switches\\ 24.9\%\end{tabular} & \begin{tabular}[c]{@{}c@{}}Routers\\ 23.7\%\end{tabular} \\ \midrule
CNN                                                                  & 73.33 / 70.78                                             & \multicolumn{1}{c|}{79.88 / 78.52}                                            & 87.77 / 88.02                                             & 74.79 / 74.19                                            & 91.24 / 95.47 / 93.31                                     & \multicolumn{1}{c|}{87.55 / \textbf{99.88} / 93.31}                                    & 84.31 / 92.23 / 88.09                                     & 92.06 / 91.69 / 91.88                                    \\
BiLSTM                                                               & 75.00 / 74.63                                             & \multicolumn{1}{c|}{79.28 / 78.31}                                            & 88.88 / 88.88                                             & 76.42 / 75.13                                            & 90.62 / 98.63 / 94.46                                     & \multicolumn{1}{c|}{91.94 / 95.82 / 93.84}                                    & 83.33 / 92.23 / 87.55                                     & 94.89 / 88.14 / 91.39                                    \\
Biglog                                                               & 88.76 / 88.68                                             & \multicolumn{1}{c|}{93.96 / 92.84}                                            & 92.38 / 91.31                                             & 82.39 / 79.33                                            & 95.40 / 98.37 / 96.86                                     & \multicolumn{1}{c|}{95.63 / 97.04 / 96.33}                                    & 90.68 / 93.58 / 92.11                                     & \textbf{95.37} / 91.16 / 93.22                                    \\
KnowLog                                                              & 91.96 / 91.05                                             & \multicolumn{1}{c|}{94.02 / 93.24}                                            & \textbf{92.39} / 90.75                                             & 83.10 / 79.15                                            & 94.99 / 97.23 / 96.10                                     & \multicolumn{1}{c|}{94.22 / 98.84 / 93.31}                                    & \textbf{94.63} / 90.38 / 92.45                                     & 93.22 / 93.97 / \textbf{93.60}                                    \\ \midrule
BERT                                                                 & 87.64 / 87.67                                             & \multicolumn{1}{c|}{91.78 / 89.72}                                            & 91.37 / 89.09                                             & 80.98 / 77.11                                            & 94.50 / 96.88 / 95.68                                     & \multicolumn{1}{c|}{95.65 / 97.48 / 96.56}                                    & 92.48 / 90.70 / 91.58                                     & 93.14 / 92.77 / 92.95                                    \\
\quad  $\mapsto Easy \: Samples$                                                        & \textit{92.06 / 91.98 }                                          & \multicolumn{1}{c|}{\textit{97.99 / 97.16}}                                            & \textit{96.59 / 95.54}                                             & \textit{86.27 / 84.40}                                            & \textit{96.66 / 99.48 / 98.05 }                                    & \multicolumn{1}{c|}{\textit{96.83 / 98.97 / 97.89}}                                    & \textit{98.16 / 95.98 / 97.06}                                     & \textit{95.87 / 96.87 / 96.37}                                    \\
\quad  $\mapsto Hard \: Samples$                                                         & \textit{76.92 / 76.51}                                            & \multicolumn{1}{c|}{\textit{75.65 / 69.88}}                                            & \textit{76.00 / 70.61}                                             & \textit{67.50 / 59.73}                                            & \textit{80.64 / 80.64 / 80.64}                                     & \multicolumn{1}{c|}{\textit{87.64 / 87.64 / 87.64}}                                    & \textit{78.16 / 77.27 / 77.71 }                                    & \textit{83.33 / 78.94 / 81.08}                                    \\ \midrule
ChatGPT                                                              & \textbf{95.55} / \textbf{95.43}                                            & \multicolumn{1}{c|}{90.82 / 92.09}                                            & 89.84 / 89.89                                             & \textbf{89.43} / \textbf{88.35}                                            & 95.69 / \textbf{98.81} / 97.23                                     & \multicolumn{1}{c|}{93.15 / 96.90 / 94.99}                                    & 85.92 / 93.91 / 89.73                                     & 90.27 / 93.17 / 91.69                                    \\ \midrule
AdaptiveLog (ICL)                                                    & 89.88 / 89.71                                             & \multicolumn{1}{c|}{92.99 / 93.19}                                            & 91.87 / 91.18                                             & 83.09 / 80.47                                            & 95.28 / \textbf{98.81} / 97.01                                     & \multicolumn{1}{c|}{95.55 / 98.37 / 96.94}                                    & 91.02 / \textbf{94.23} / \textbf{92.59}                                     & 92.15 / 94.37 / 93.25                                    \\
\quad  $\mapsto Hard \: Samples$                                                        & \textit{84.61 / 84.03}                                             & \multicolumn{1}{c|}{\textit{80.00 / 82.04}}                                            & \textit{78.00 / 78.01}                                             & \textit{75.00 / 71.93 }                                           & \textit{87.12 / 94.62 / 90.72 }                                    & \multicolumn{1}{c|}{\textit{87.50 / 94.38 / 90.81}}                                    & \textit{75.96 / 89.77 / 82.29}                                     & \textit{80.32 / 85.96 / 83.05}                                    \\ \midrule
AdaptiveLog (ECR)                                                    & 91.57 / 91.36                                           & \multicolumn{1}{c|}{\textbf{94.44 }/ \textbf{94.24}}                                            & 91.87 / \textbf{92.16}                                             & 85.91 / 85.50                                            & \textbf{96.23} / 98.51 / \textbf{97.36}                                     & \multicolumn{1}{c|}{\textbf{96.09} / 98.07 / \textbf{97.07}}                                    & 91.25 / 93.58 / 92.40                                     & 91.18 / \textbf{95.58} / 93.33                                    \\
\quad  $\mapsto Hard \: Samples$                                                        & \textit{90.38 / 90.08}                                             & \multicolumn{1}{c|}{\textit{85.21 / 86.42}}                                            & \textit{78.00 / 75.61}                                             & \textit{85.00 / 86.47}                                            & \textit{93.47 / 92.47 / 92.97  }                                   & \multicolumn{1}{c|}{\textit{91.11 / 92.13 / 91.62}}                                    & \textit{76.23 / 87.50 / 81.48}                                     & \textit{77.61 / 91.22 / 83.87}                                    \\ \bottomrule[1.5pt]
\end{tabular}%
}
\end{table}

\begin{table}[]
\centering
\caption{Results on Log and Description Semantic Matching and Log and Possible Cause Ranking.}
\label{tab:main_exp3}
\resizebox{0.8\columnwidth}{!}{%
\begin{tabular}{lcccc|cc}
\toprule[1.5pt]
Models                                                                & \multicolumn{4}{c|}{LDSM (Accuracy / Weighted-F1)}                                                                                                                                                                                                                & \multicolumn{2}{c}{LPCR (Precision @ 1/ MRR)}                                                                        \\ \midrule
                                                                      & \multicolumn{2}{c|}{Cisco}                                                                                                                & \multicolumn{2}{c|}{Huawei}                                                                                           & \multicolumn{2}{c}{Huawei}                                                                                           \\
\begin{tabular}[c]{@{}l@{}}Percentage of \\ hard samples\end{tabular} & \begin{tabular}[c]{@{}c@{}}Switches\\ 28.6\%\end{tabular} & \multicolumn{1}{c|}{\begin{tabular}[c]{@{}c@{}}Routers\\ 27.4\%\end{tabular}} & \begin{tabular}[c]{@{}c@{}}Switches\\ 21.3\%\end{tabular} & \begin{tabular}[c]{@{}c@{}}Routers\\  29.3\%\end{tabular} & \begin{tabular}[c]{@{}c@{}}Switches\\ 30.0\%\end{tabular} & \begin{tabular}[c]{@{}c@{}}Routers\\ 34.9\%\end{tabular} \\ \midrule
CNN                                                                   & 72.48 / 72.42                                             & \multicolumn{1}{c|}{73.97 / 73.90}                                            & 85.69 / 85.68                                             & 82.63 / 82.62                                             & 64.87 / 79.20                                             & 53.88 / 72.67                                            \\
BiLSTM                                                                & 73.61 / 73.57                                             & \multicolumn{1}{c|}{75.59 / 75.59}                                            & 85.81 / 85.79                                             & 82.57 / 82.57                                             & 65.78 / 79.61                                             & 56.11 / 74.92                                            \\
Biglog                                                                & 89.07 / 85.94                                             & \multicolumn{1}{c|}{88.10 / 88.10}                                            & 95.21 / 95.21                                             & 93.31 / 93.30                                             & 89.47 / 94.07                                             & 86.52 / 92.74                                            \\
KnowLog                                                               & 89.50 / 89.48                                             & \multicolumn{1}{c|}{90.15 / 90.14}                                            & 95.99 / 95.99                                             & 94.13 / 94.13                                             & 91.09 / 94.87                                             & 87.91 / 93.24                                            \\ \midrule
BERT                                                                  & 85.95 / 85.94                                             & \multicolumn{1}{c|}{88.01 / 87.99}                                            & 94.23 / 94.23                                             & 89.15 / 89.13                                             & 87.34 / 92.63                                             & 86.11 / 92.16                                            \\
\quad  $\mapsto Easy \: Samples$                                                          & \textit{93.17 / 93.17 }                                            & \multicolumn{1}{c|}{\textit{94.36 / 94.36}}                                            & \textit{98.12 / 98.12}                                             & \textit{96.59 / 96.59 }                                            & \textit{94.36 / 96.43 }                                            & \textit{89.12 / 93.65}                                            \\
\quad  $\mapsto Hard \: Samples$                                                          & \textit{67.88 / 67.78}                                             & \multicolumn{1}{c|}{\textit{71.13 / 70.93}}                                            & \textit{79.88 / 79.87 }                                            & \textit{71.16 / 71.02}                                             & \textit{70.94 / 83.69}                                             & \textit{80.47 / 88.79}                                            \\ \midrule
ChatGPT                                                               & 81.07 / 80.81                                                         & \multicolumn{1}{c|}{80.39 / 80.08}                                                        & 90.35 / 90.31                                             & 87.45 / 87.35                                             & 90.28 / 94.78                                             & 87.77 / 93.45                                            \\ \midrule
AdaptiveLog (ICL)                                                     & 86.57 / 86.51                                             & \multicolumn{1}{c|}{90.32 / 90.28}                                            & 94.76 / 94.76                                             & 92.00 / 92.00                                             & 91.90 / 95.22                                             & 87.91 / 93.26                                            \\
\quad  $\mapsto Hard \: Samples$                                                          & \textit{70.07 / 68.83}                                             & \multicolumn{1}{c|}{\textit{73.91 / 73.03}}                                            & \textit{82.37 / 82.06 }                                            & \textit{80.89 / 80.77 }                                            & \textit{86.14 / 92.39 }                                            & \textit{85.65 / 91.95  }                                     \\ \midrule
AdaptiveLog (ECR)                                                     & \textbf{90.03} / \textbf{90.03}                                             & \multicolumn{1}{c|}{\textbf{90.91} / \textbf{90.91}}                                            & \textbf{96.72} / \textbf{96.72}                                             & \textbf{95.39} / \textbf{95.39}                                             & \textbf{94.02} / \textbf{96.40}                                             & \textbf{91.25} / \textbf{95.13}                                            \\
\quad  $\mapsto Hard \: Samples$                                                          & \textit{82.18 / 82.18}                                             & \multicolumn{1}{c|}{\textit{81.73 / 81.73}}                                            & \textit{91.57 / 91.55}                                             & \textit{92.50 / 92.53}                                             & \textit{93.24 / 96.32}                                             & \textit{95.21 / 97.31}                                            \\ \bottomrule[1.5pt]
\end{tabular}%
}
\end{table}

\subsubsection{\textbf{RQ1: How effective is AdaptiveLog compared with the current mainstream methods on downstream tasks?}} 
This section compares the AdaptiveLog with other state-of-the-art log analysis methods on six downstream tasks. In addition, 
AdaptiveLog divides the simple and hard samples based on the probability of uncertainty. To more intuitively analyze the model's performance on different uncertain samples, we also give results on simple and hard samples of the SLM respectively. The experiment results are shown in Table \ref{tab:main_exp1} - \ref{tab:main_exp3}.

From the results, it is clear that AdaptiveLog outperforms all baselines among software system and network device logs, demonstrating not only the effectiveness of AdaptiveLog for automated log analysis but also the ability to generalize across different types of logs. \textbf{(1)} Firstly, compared with traditional deep-learning methods and existing pre-trained models, especially KnowLog, which is the current state-of-the-art model on log analysis tasks and even better than the LLM on the specific log analysis task due to its abundant domain knowledge.
Its advantage lies in harnessing the powerful knowledge and reasoning capabilities of LLMs to analyze logs beyond the capacity of SLMs, thereby compensating for their inherent limitations.
\textbf{(2)} Secondly,  in comparison to ChatGPT, AdaptiveLog not only demonstrates superior performance but also exhibits significant cost advantages, requiring less than 30\% of the average cost. This is because after fine-tuning, the SLM can better capture task-specific details and features for most logs, and LLMs have the advantage of powerful reasoning and learning from a few examples to analyze those hard logs with complex intricate patterns. Hence, on feature-simple samples, the LLM may not necessarily outperform a specific SLM, and this hybrid strategy effectively amalgamates the strengths of both models.
\textbf{(3)} Thirdly, the evaluation of the SLM's (BERT) performance on simple and hard samples reveals a notable discrepancy, where the results on simple samples excel far beyond the SLM's overall performance, and the results of hard samples are significantly weaker than the overall performance. This indicates the efficacy of our proposed uncertainty estimation strategy in discerning the SLM's capabilities. In particular, hard sample performance substantially reduces the overall performance of the SLM, which is beyond the ability of the SLM. After analysis by the LLM, the performance of the hard samples is also significantly improved compared to the SLM. This underscores the advantage of LLMs in handling complex logs, thereby facilitating significant cost savings through improved efficiency.

In addition, we also compare our proposed ECR strategy with ICL. From these results, ECR outperforms the standard ICL strategy, which indicates that the LLM can learn and benefit from error-prone cases. Compared with the standard ICL, ECR provides the reasoning process and potential pitfalls of error-prone cases for the LLM, and this strategy not only improves the LLM's performance on challenging log samples but also serves as a valuable learning mechanism, guiding the model to avoid similar errors in future analyses. The ECR strategy contributes to the robustness and adaptability of AdaptiveLog in handling complex log data.

Consequently, we can conclude that the comprehensive analysis of AdaptiveLog underscores its strengths in complex log analysis tasks. In addition, when compared to entirely analyzing logs with LLMs, AdaptiveLog not only demonstrates superior performance but also notably mitigates the associated costs linked with querying LLMs.

    

\begin{figure}[]
\centering
   \includegraphics[width=0.84\linewidth]{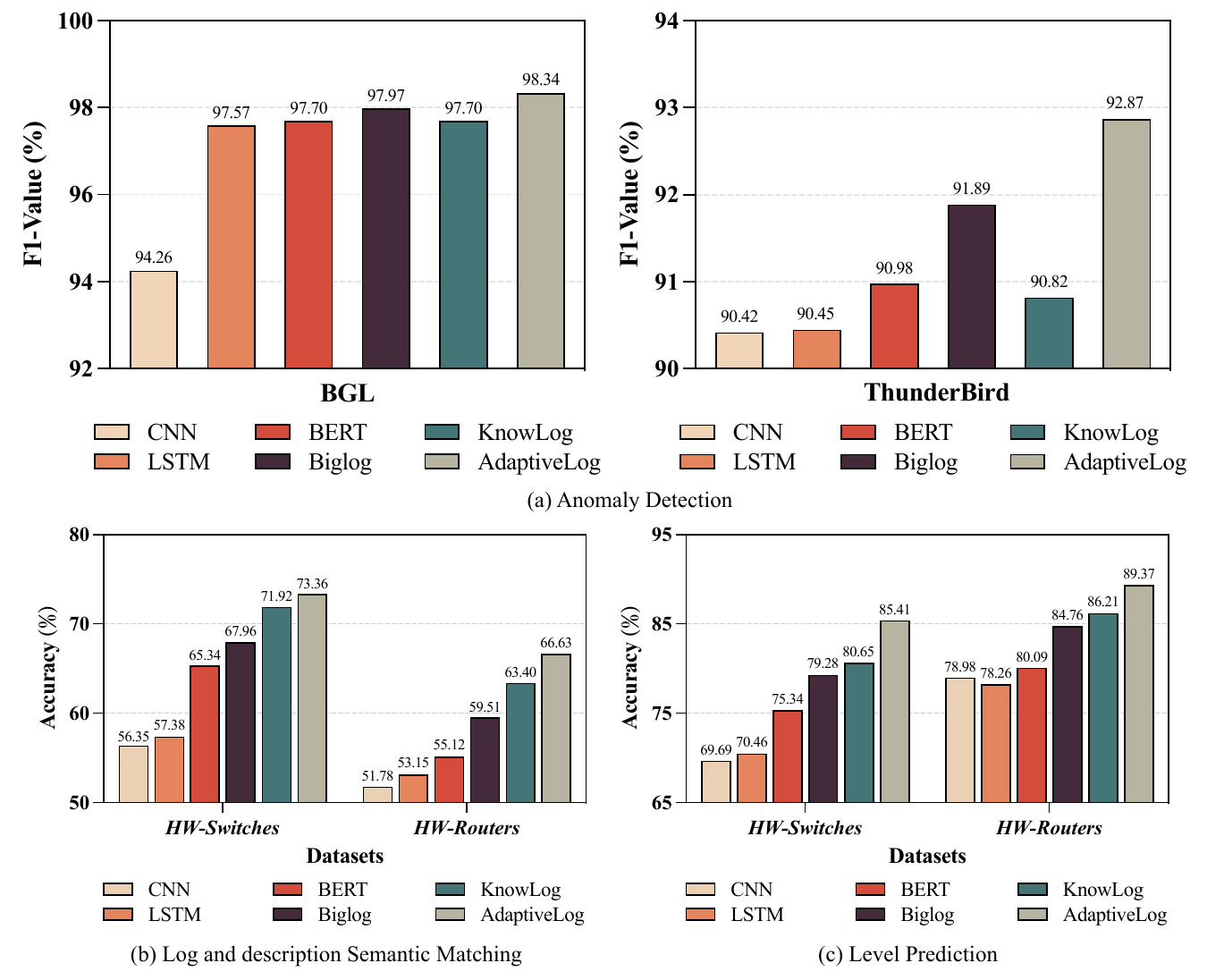}
   \caption{Results in the low-resource scenario.}
   \label{fig:low-resource result}
\end{figure}

\subsubsection{\textbf{RQ2: How effective is AdaptiveLog in the low-resource scenario?}}
In real-world log analysis scenarios, due to the massive size of logs, it's usually difficult to obtain sufficient labeled data \cite{le2022log}, which is a challenge for data-driven log analysis methods. 
To verify the effectiveness of AdaptiveLog in this low-resource scenario, we conduct experiments on Anomaly detection, Log and Description Semantic Matching, and  Level Prediction three tasks. Following these existing works, we assume that only 10\% of the labeled data is available in the training set for each task, which means that the smaller pre-trained model is fine-tuned on only 10\% of training data. 

The experimental results are shown in Fig. \ref{fig:low-resource result}, which shows that AdaptiveLog performs much better than other baselines. Firstly, traditional deep learning models CNN and BiLSTM exhibit poor performance in low-resource scenarios. This is indicative of their struggle to extract meaningful features from limited samples for effective log analysis. The complex patterns present in log data require sophisticated feature extraction mechanisms, a challenge that traditional deep models find difficult to overcome without sufficient data.
Secondly, models like Biglog and KnowLog, while showing some improvement over models like BERT, still fall short of AdaptiveLog's performance. The key advantage of AdaptiveLog lies in its ability to leverage the strengths of LLMs in handling complex samples. By providing complex samples to the LLM, the framework enables the LLM to utilize its embedded knowledge for effective reasoning in low-resource settings. This reasonable utilization of LLMs' capabilities significantly enhances the framework's performance.
In addition, we also count the proportion of hard samples on the Huawei-Switches dataset for both tasks, which is 32.6\% and 40.3\% respectively. Compared to the full sample fine-tuned BERT, the proportion of hard samples increased by 11.3\% and 15.4\%. This also illustrates the need for the LLM to analyze more samples when the SLM is weaker, which also inspires us that we can leverage the capabilities of LLMs better to compensate for the shortcomings of the SLM in the low resource scenario,  where the advantages of LLMs in handling complex log data become more evident and essential for effective analysis.

\begin{figure}[]
    \centering
    \subfloat[Level Prediction (Different Vendor) ]{\label{fig: system type forgetting}\includegraphics[width=0.33\linewidth]{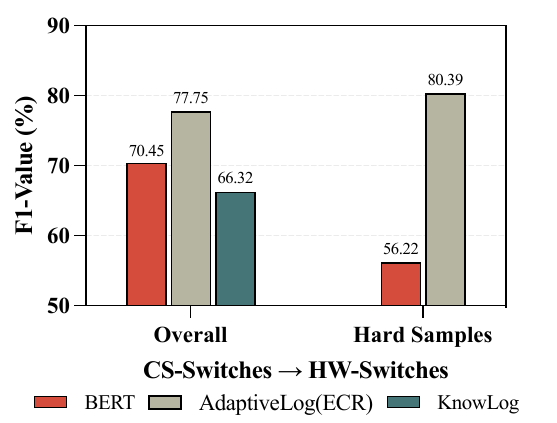}}
    \subfloat[LDSM (Different Vendor)]{\label{fig: system type zero}\includegraphics[width=0.33\linewidth]{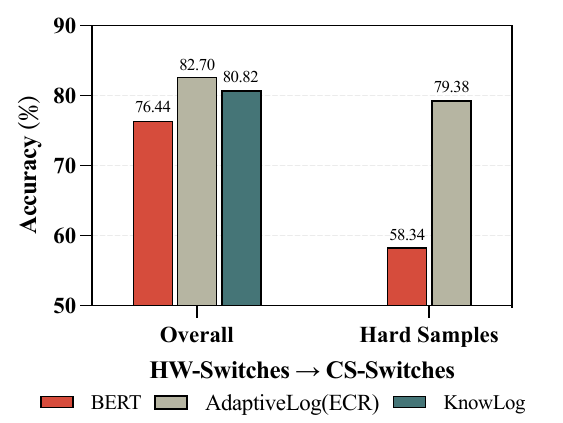}}
    \subfloat[LPCR (Different Device)]{\label{fig: system type zero}\includegraphics[width=0.33\linewidth]{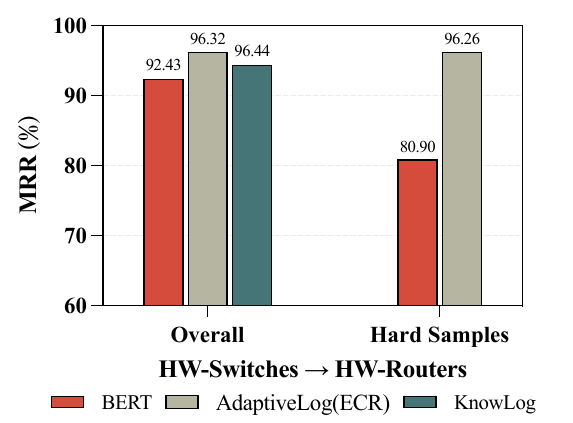}}

    \subfloat[Level Prediction (Different Vendor)]{\label{fig: system type forgetting}\includegraphics[width=0.33\linewidth]{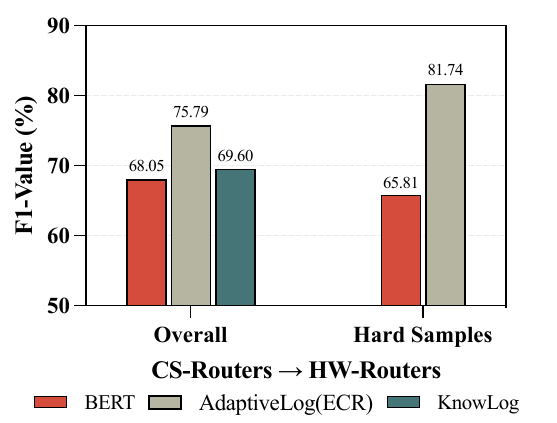}}
    \subfloat[LDSM (Different Vendor)]{\label{fig: system type zero}\includegraphics[width=0.33\linewidth]{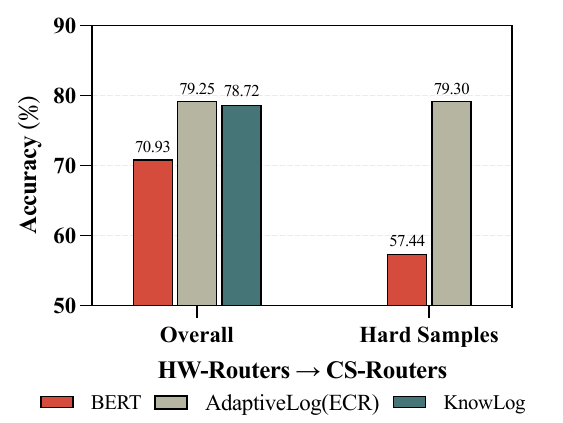}}
    \subfloat[LPCR (Different Device)]{\label{fig: system type zero}\includegraphics[width=0.33\linewidth]{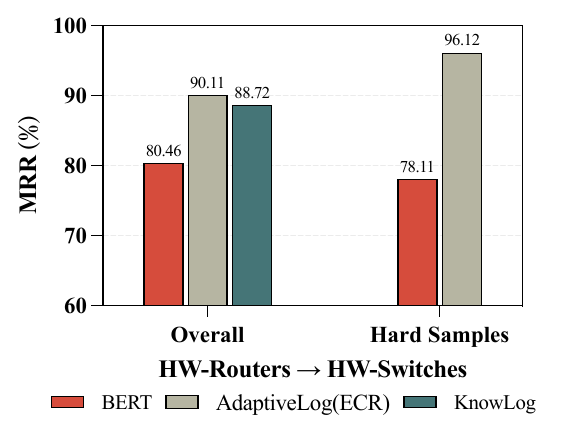}}
    \caption{Results in the transfer learning scenario. Left side of $\rightarrow$ indicates the source dataset for training and right side indicates the target dataset for testing. }
    \label{fig:transfer_exp}
\end{figure}

We can conclude that AdaptiveLog achieves outstanding performance in low-resource scenarios. By recognizing the importance of leveraging LLMs in compensating for SLM weaknesses in low-resource settings, AdaptiveLog demonstrates outstanding performance and underscores the significant advantages of our proposed framework in challenging log analysis scenarios with limited resources.

\subsubsection{\textbf{RQ3: How effective is AdaptiveLog in the transfer learning scenario?}}
In the context of transfer learning for log analysis, where models need to adapt to different vendor logs due to variations in syntax \cite{chen2020logtransfer}, the effectiveness of AdaptiveLog in transfer scenarios becomes a key focus. Deploying different models for specific tasks based on vendor logs incurs the challenge of requiring annotated data from each vendor, resulting in operational inefficiencies. The expectation is to train one single model on annotated data from one vendor and successfully transfer its learning to analyze logs from different vendors, showcasing the model's transfer learning capabilities. To assess the transfer ability of models across different vendor logs, we conduct experiments in Level Prediction, Log and Semantic Description, and Log and Possible Cause Ranking three tasks. Specifically, the SLM is trained using a training dataset from a source vendor or device and subsequently evaluated on a separate testing dataset from a target vendor or device.

The experimental results are shown in Fig. \ref{fig:transfer_exp}, it can be seen that Adaptivelog significantly enhances the performance of the SLM in transfer learning. This growth comes from the collaborative reasoning of LLM, and for the SLM of hard samples, the ECR reasoning strategy grows up to 24.17\% F1-value on the Level Prediction task. The successful evaluation of AdaptiveLog on different tasks across different vendor and device logs underscores the framework's transfer capability. By training on one vendor's dataset and seamlessly applying the learned knowledge to logs from another vendor, AdaptiveLog showcases its versatility and adaptability in handling diverse log data sources. The transfer ability of AdaptiveLog offers operational advantages by minimizing the complexities associated with training and deploying separate models for distinct vendors. By enabling seamless knowledge transfer between vendor logs, the framework promotes operational efficiency, reduces the need for extensive labeled data, and facilitates a more unified and streamlined approach to log analysis across different vendors.

In conclusion, AdaptiveLog proves to be highly effective in the transfer scenario, showcasing robust transfer learning capabilities in analyzing logs from different vendors. This ability to generalize across vendor logs not only streamlines the log analysis process but also enhances scalability and applicability in diverse log analysis settings, highlighting the significant advantages of transfer learning with AdaptiveLog in the transfer scenarios.

\begin{table*}[]
\caption{Results on the collaboration with different SLMs, including two traditional deep-learning models and three different structures of SLMs .}
\label{tab:others}
\resizebox{\textwidth}{!}{%
\begin{tabular}{c|lcc|c|lcc}
\toprule[1.5pt]
Style                                                                                   & Models                                         & Huawei-Switches        & Huawei-Routers         & Style                                                                             & Models                                         & Huawei-Switches        & Huawei-Routers         \\ \midrule
\multirow{5}{*}{\begin{tabular}[c]{@{}c@{}}Encoder-\\ only\\ SLM\end{tabular}}          & Biglog                                         & 95.21 / 95.21          & 93.31 / 93.31          & \multirow{5}{*}{\begin{tabular}[c]{@{}c@{}}Encoder-\\ only\\ SLM\end{tabular}}    & KnowLog                                        & 95.99 / 95.99          & 94.13 / 94.13          \\
                                                                                        & \quad $\mapsto Easy \: Samples$ & 95.86 / 95.86          & 99.27 / 99.27          &                                                                                   & \quad $\mapsto Easy \: Samples$ & 98.56 / 98.56          & 97.00 / 97.00          \\
                                                                                        & \quad $\mapsto Hard \: Samples$ & 74.91 / 74.34          & 80.51 / 80.42          &                                                                                   & \quad $\mapsto Hard \: Samples$ & 80.79 / 80.77          & 80.52 / 80.43          \\  \cmidrule{2-4} \cmidrule{6-8} 
                                                                                        & AdaptiveLog (ECR)                              & \textbf{96.89} / \textbf{96.89}          & \textbf{96.34} / \textbf{96.34}          &                                                                                   & AdaptiveLog (ECR)                              & \textbf{96.34} / \textbf{96.34}          & \textbf{96.16} / \textbf{96.16}          \\
                                                                                        & \quad $\mapsto Hard \: Samples$ & 89.09 / 88.92          & 90.34 / 90.32          &                                                                                   & \quad $\mapsto Hard \: Samples$ & 87.85 / 87.77          & 90.13 / 90.11          \\ \midrule
\multirow{10}{*}{\begin{tabular}[c]{@{}c@{}}Deep Learning-\\ based Models\end{tabular}} & CNN                                            & 84.51 / 84.43          & 82.35 / 82.22          & \multirow{5}{*}{\begin{tabular}[c]{@{}c@{}}Decoder-\\ only\\ SLM\end{tabular}}    & Llama-3.2-3B-Instruct                          & 75.15 / 75.13          & 75.01 / 75.02          \\
                                                                                        & \quad $\mapsto Easy \: Samples$ & 89.50 / 89.44          & 86.44 / 86.32          &                                                                                   & \quad $\mapsto Easy \: Samples$ & 88.79 / 88.76          & 89.31 / 89.30          \\
                                                                                        & \quad $\mapsto Hard \: Samples$ & 72.15 / 72.05          & 70.21 / 70.12          &                                                                                   & \quad $\mapsto Hard \: Samples$ & 43.87 / 43.85          & 45.08 / 45.00          \\ \cmidrule{2-4} \cmidrule{6-8} 
                                                                                        & AdaptiveLog (ECR)                              & \textbf{89.04 / 88.98} & \textbf{85.28 / 85.28} &                                                                                   & AdaptiveLog (ECR)                              & \textbf{88.27 / 88.23} & \textbf{85.91 / 85.90} \\
                                                                                        & \quad $\mapsto Hard \: Samples$ & 87.92 / 87.86          & 80.15 / 80.65          &                                                                                   & \quad $\mapsto Hard \: Samples$ & 87.07 / 87.00          & 78.81 / 78.03          \\ \cmidrule{2-8} 
                                                                                        & BiLSTM                                         & 84.71 / 84.61          & 82.57 / 82.42          & \multirow{5}{*}{\begin{tabular}[c]{@{}c@{}}Encoder-\\ Decoder\\ SLM\end{tabular}} & ChatGLM-6B                                     & 70.33 / 82.58          & 72.05 / 83.75          \\
                                                                                        & \quad $\mapsto Easy \: Samples$ & 91.83 / 91.74          & 86.37 / 86.18          &                                                                                   & \quad $\mapsto Easy \: Samples$ & 84.07 / 91.34          & 83.71 / 91.13          \\
                                                                                        & \quad $\mapsto Hard \: Samples$ & 72.30 / 72.41          & 74.66 / 74.61          &                                                                                   & \quad $\mapsto Hard \: Samples$ & 39.70 / 56.84          & 42.55 / 59.70          \\ \cmidrule{2-4} \cmidrule{6-8} 
                                                                                        & AdaptiveLog (ECR)                              & \textbf{89.90 / 89.84} & \textbf{84.98 / 84.97} &                                                                                   & AdaptiveLog (ECR)                              & \textbf{86.88 / 85.28} & \textbf{81.64 / 82.87} \\
                                                                                        & \quad $\mapsto Hard \: Samples$ & 86.54 / 86.59          & 82.09 / 81.72          &                                                                                   & \quad $\mapsto Hard \: Samples$ & 87.94 / 87.99          & 76.40 / 75.08          \\ \bottomrule[1.5pt]
\end{tabular}%
}
\end{table*}

\subsubsection{\textbf{RQ4: How effective is AdaptiveLog combined with different SLMs?}}
Considering that there will be different SLMs in the real scenario, to verify the flexibility of AdaptiveLog, we conduct experiments on deep learning models and other SLMs. The key for AdaptiveLog to combine with other SLMs is that Bayesian inference is effective on these SLMs. Specifically, apart from BERT, 
we exploit traditional deep-learning models (CNN and BiLSTM) and different structures of SLMs, including encoder-only (Biglog and KnowLog), decoder-only (\textit{Llama-3.2-3B-Instruct}), and encoder-decoder models (\textit{ChatGLM-6B}), as SLMs and conduct experiments on the Log and Description Semantic Matching task. For generative SLMs, we calculate the average probability of all output tokens as the prediction probability to estimate uncertainty.

The experimental results are shown in Table \ref{tab:others}, we find that both models can outperform the original model when combined with the LLM, which implies that AdaptiveLog is effective in combination with different SLMs. 
In addition, from the perspective of growth, we also find that the weaker the performance of the SLM, the more the final result is improved when combined with the LLM. This implies that Bayesian inference is effective in filtering out hard samples to give to the LLM, and also validates the effectiveness of Bayesian inference in estimating the uncertainty of different SLMs.
Due to pre-training on log corpus, Biglog and
KnowLog outperforms BERT and it has been shown that as the capabilities of the SLM increase, the final performance of AdaptiveLog usually improves as well. We choose BERT as the SLM in our experiments due to its easier accessibility and wider application compared to Biglog and KnowLog. Despite combining with the weaker SLM, AdaptiveLog still outperforms other state-of-the-art methods. In fact, AdaptiveLog can be effectively combined with other SLMs as well.

In conclusion, AdaptiveLog can combine with different SLMs well, and the more powerful the SLM, the better performance AdaptiveLog ultimately achieves. We encourage users to use the more powerful SLM to combine with AdaptiveLog, this allows for superior performance.

\begin{figure}
    \centering
    \subfloat[LDSM (HW-Switches)]{\label{fig: Hw-Switches}\includegraphics[width=0.5\linewidth]{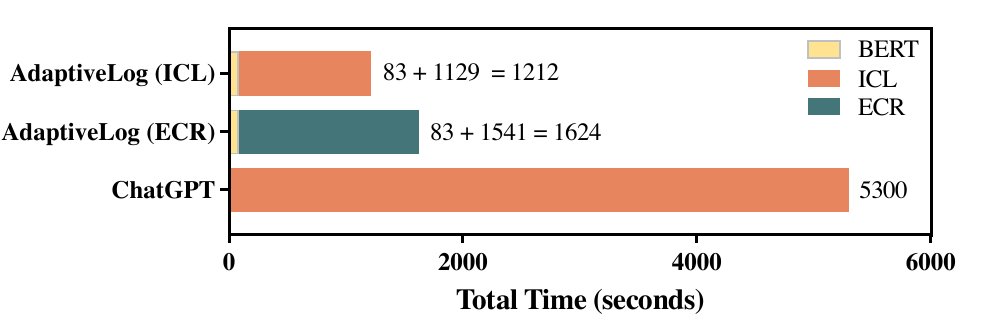}}
    \subfloat[LPCR (HW-Switches)]{\label{fig: system type zero}\includegraphics[width=0.5\linewidth]{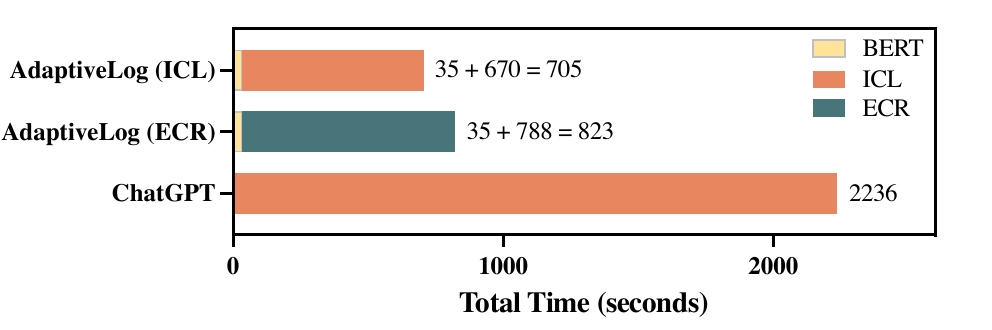}}
    \caption{Efficieny of AdaptiveLog and ChatGPT on different tasks.}
    \label{fig:total_time_costs}
\end{figure}

\subsubsection{\textbf{RQ5: How efficient is AdaptiveLog in the automated log analysis scenario?}} In this section, our primary focus is on assessing the efficiency of the execution process within AdaptiveLog. More specifically, we record the execution times for AdaptiveLog including uncertainty estimation time and ECR (or ICL) time. We take BERT as the SLM and deploy it on one NVIDIA RTX 3090 GPU, and then we count the total running time of Adaptivelog during prediction on LDSM and LPCR tasks, where the SLM needs to perform 10 observations to calculate the uncertainty and hard samples are analyzed by the ChatGPT.  Additionally, the runtime of ICL-based ChatGPT on all test samples is also recorded for comparison.

According to the results in Fig. \ref{fig:total_time_costs}, we can see that AdaptiveLog demonstrates remarkable efficiency compared to ChatGPT. Notably, the uncertainty estimation time for SLMs is exceptionally fast, significantly outperforming the querying of LLMs. The average time spent on uncertainty estimation is merely 4.68\% of the total AdaptiveLog runtime, highlighting the efficiency of this process within the framework.
Secondly, the ECR-based reasoning strategy, while slightly slower than the ICL-based strategy on hard samples, showcases a meticulous approach in analyzing the reasoning process of complex log samples. Despite this, the time cost associated with the ECR-based strategy remains acceptable, especially considering the performance improvement it offers. Even with this slight time overhead, AdaptiveLog still significantly outperforms ChatGPT, which conducts full-sample inference. On average, AdaptiveLog saves 66.27\% of the runtime compared to ChatGPT, emphasizing its superior efficiency in automated log analysis tasks.

\begin{figure}[]
    \centering
    \subfloat[LDSM (HW-Switches)]{\label{fig: Hw-Switches}\includegraphics[width=0.45\linewidth]{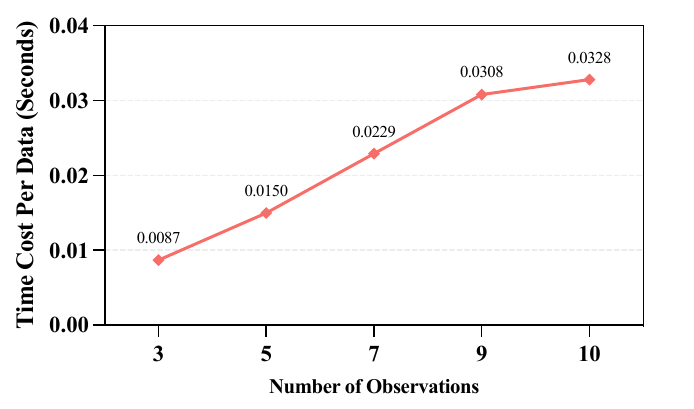}}
    \subfloat[LPCR (HW-Switches)]{\label{fig: system type zero}\includegraphics[width=0.45\linewidth]{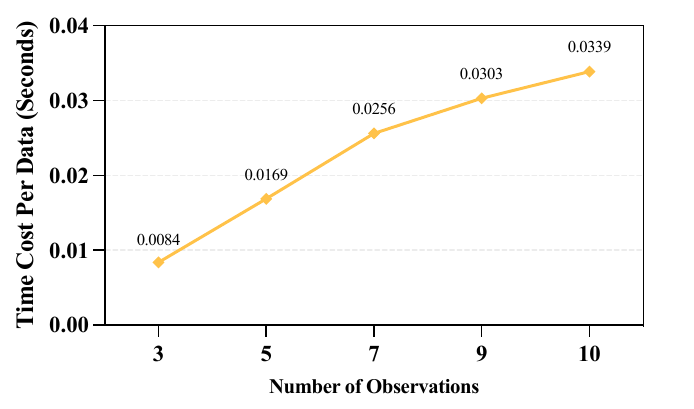}}
    \caption{Time with different number of observations on LDSM and LPCR tasks.}
    \label{fig: Time_uncertain_cost}
\end{figure}

\begin{figure}[]
    \centering
    \subfloat[LDSM (HW-Switches)]{\label{fig: Hw-Switches}\includegraphics[width=0.45\linewidth]{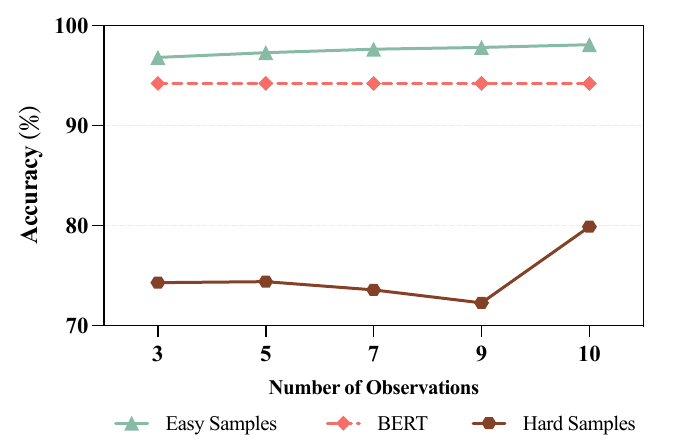}}
    \subfloat[LPCR (HW-Switches)]{\label{fig: system type zero}\includegraphics[width=0.45\linewidth]{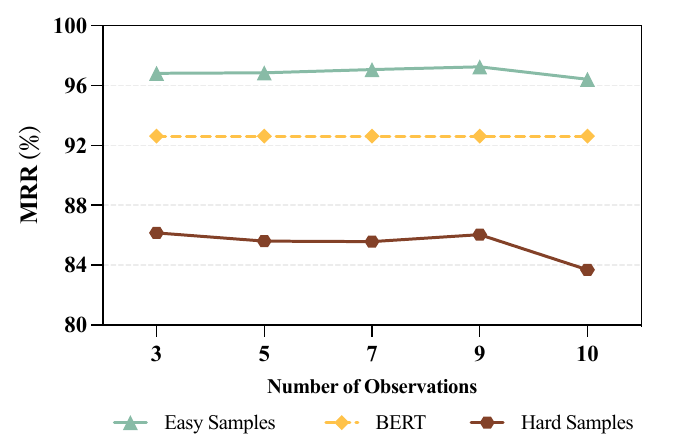}}
    \caption{Results with different number of observations on LDSM and LPCR tasks.}
    \label{fig:result_observations}
\end{figure}

In addition, to determine the estimation time of uncertainty with respect to the number of observations, we also calculate the times corresponding to different numbers of observations on the LDSM and LPCR tasks. As shown in Fig. \ref{fig: Time_uncertain_cost}, we can find that uncertainty estimation time increases linearly with the number of observations. This means that uncertainty estimation for SLMs does not introduce a huge time overhead. 
We also analyze the effect of different number of observations on the uncertainty estimation, and the results in Fig. \ref{fig:result_observations} show that 3 observations can also effectively determine the SLM's uncertainty, spending only an average of 0.0086 seconds per data, which can also reduce the uncertainty estimation overhead. Empirically, the more observations are made, the more stable the results become, and the user can set the number of observations according to their needs.
This efficiency, combined with the framework's superior performance and adaptability, positions AdaptiveLog as a highly efficient and effective solution for automated log analysis, offering enhanced productivity and performance in real-world log analysis scenarios.

In conclusion, the efficiency analysis of AdaptiveLog in automated log analysis tasks highlights its exceptional performance and operational advantages. The framework's ability to conduct fast and accurate uncertainty estimation with SLMs, coupled with the strategic reasoning process facilitated by the ECR-based strategy, underscores its efficiency in log analysis tasks. Despite the slightly increased time cost associated with the ECR approach on hard samples, the overall efficiency gains of AdaptiveLog are substantial, showcasing significant time savings compared to LLMs.

\subsubsection{\textbf{RQ6: How do different ECR strategies affect effectiveness?}}
This section analyzes the impact of different configurations on ECR from the following two aspects: (1) number of error cases, (2) strategy of case selection.
By default, ECR selects 5 most similar error cases as prompt examples in each inference. Analyzing the impact of different configurations on ECR aims to allow users to choose the optimal configuration when analyzing their logs using AdaptiveLog. 

\begin{figure}[]
    \centering
    \subfloat[Results only on hard Samples (HW-Switches)]{\label{fig: system type forgetting}\includegraphics[width=0.45\linewidth]{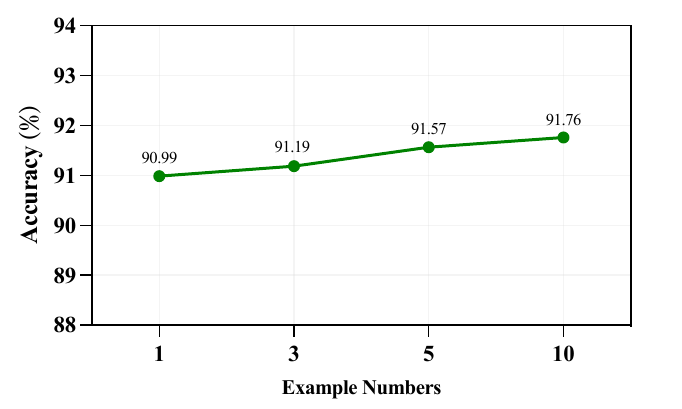}}
    \subfloat[Results only on hard Samples (HW-Routers)]{\label{fig: system type zero}\includegraphics[width=0.45\linewidth]{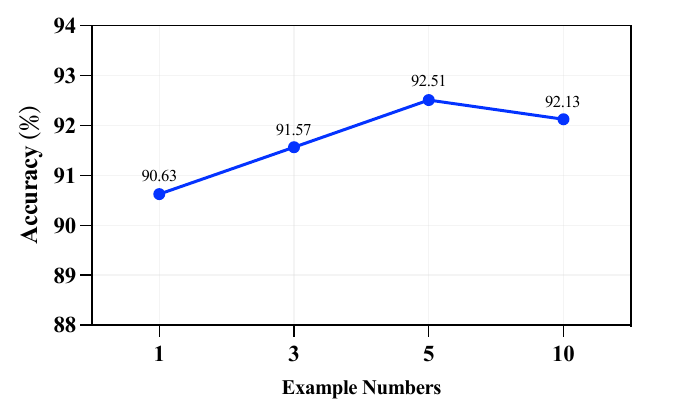}}
    \caption{Results of different error case numbers on the LDSM.}
    \label{fig:example_numbers}
\end{figure}

\begin{figure}[]
    \centering
    \subfloat[Accuracy only on Hard samples]{\label{fig: system type forgetting}\includegraphics[width=0.45\linewidth]{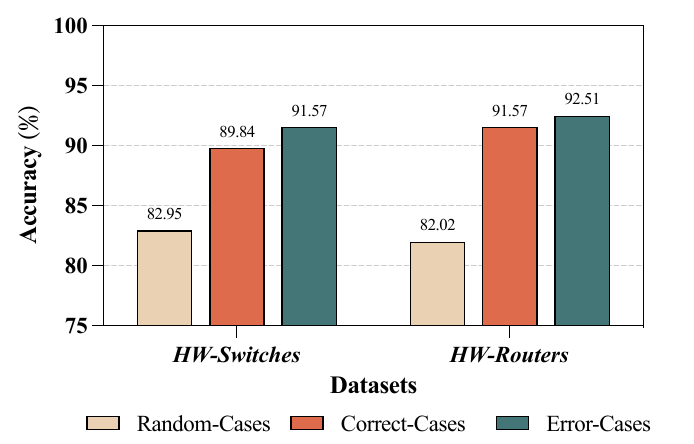}}
    \subfloat[Weighted F1  only on Hard samples]{\label{fig: system type zero}\includegraphics[width=0.45\linewidth]{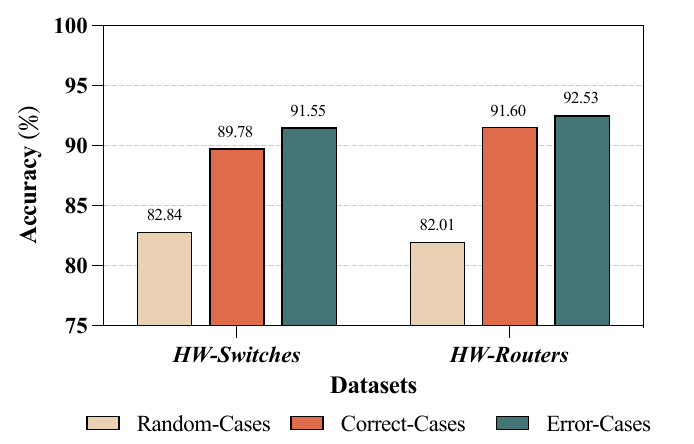}}
    \caption{Results on different case selection strategies.}
    \label{fig:case_strategy}
\end{figure}

\textbf{(1) Error Case Numbers.} To investigate the impact of error
case numbers in the prompt, we perform comparative analysis on the LDSM task by considering four different case numbers: 1, 3, 5, and 10. The experiment results are shown in Fig. \ref{fig:example_numbers}. From the results, we find a significant upward trend in model performance when the number of error cases is below 5. When the number exceeds 5, the upward trend of the model's performance is slow and even declines. This suggests that more cases are not better, and an appropriate number of error cases helps the model to learn experiences from the case. Instead, the increased number of error cases results in slower inference efficiency and increases the cost of the LLM.
Hence, we set the number of error cases to 5 in our experiments.

\textbf{(2) Case Selection Strategy.} To analyze the effectiveness of similar error-prone cases for improving the reasoning of LLM on log analysis, we also consider two different case selection strategies to compare with similar error-prone cases: random error cases and similar correct cases. For random error cases, we randomly select 5 error-prone samples from the error-prone database $\mathcal{D}$ as prompt examples, and for similar correct cases, we select 5 similar correct samples of the SLM from the validation set. We conduct experiments on the LDSM task and the results of the LLM on hard samples are shown in Fig. \ref{fig:case_strategy}.

From the result, we can find that similar error-prone cases achieve the best performance, suggesting that the LLM can benefit from error-prone cases, which facilitates the reasoning of the given logs. Compared with random error cases, the results for random error cases are significantly weaker than the other two strategies, which suggests that only similar cases can contribute to the reasoning of the LLM. Since pseudo-error cases cannot effectively capture common patterns, they can instead degrade the performance of the LLM. 
When compared with similar correct cases, similar error cases still have the advantage, which suggests that the potential pitfalls and the reasoning process of these error-prone cases are more conducive for the LLM to learn and benefit from these mistakes and thus avoid the same errors.
In conclusion, choosing suitable examples is important for the LLM. Similar error cases are more conducive to improving the reasoning of LLM on logs. Hence, we encourage users to collect error-prone cases to prompt the LLM, which can contribute to more experiences that can be referenced for the LLM.

\subsection{Ablation Studies}

\subsubsection{Component Ablation Studies}
\begin{table}[]
    \centering
    \caption{Ablation studies on ECR. We report the results of hard samples on the LDSM.}
    \label{tab:ab-ecr}
    \resizebox{0.5\textwidth}{!}{%
    \begin{tabular}{lcc}
    \toprule[1.5pt]
    Strategy       & Huawei-Switches & Huawei-Routers \\ \midrule
    ICL            & 82.37 / 82.06   & 80.89 / 80.77  \\ \midrule
    ECR            & \textbf{91.57} / \textbf{91.55}   & \textbf{92.50} / \textbf{92.53}  \\
    ECR w/o Reason & 87.35 / 87.19   & 88.20 / 88.20  \\ \bottomrule[1.5pt]
    \end{tabular}%
    }
\end{table}

\begin{figure}
    \centering
    \subfloat[Complete results on the LDSM.]{\label{fig: Hw-Switches}\includegraphics[width=0.45\linewidth]{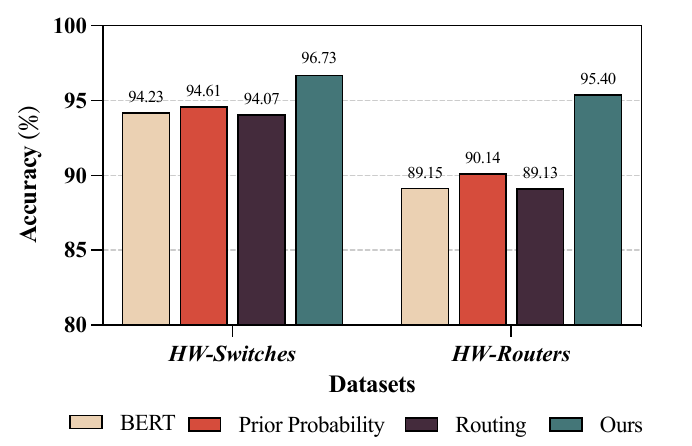}}
    \subfloat[Results only on hard samples.]{\label{fig: system type zero}\includegraphics[width=0.45\linewidth]{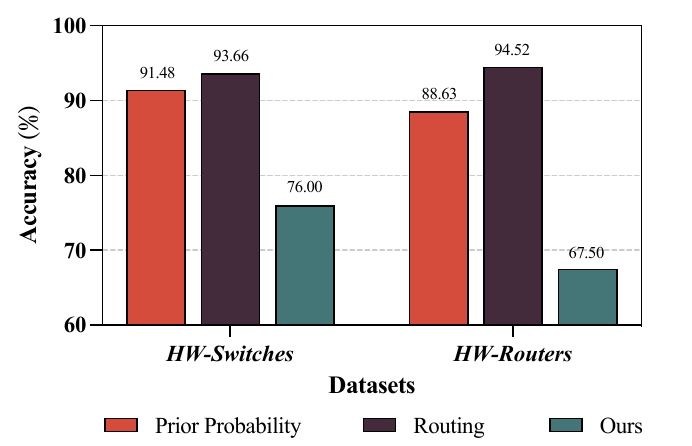}}
    \caption{Ablation studies on Bayesian inference.}
    \label{fig:Ablation}
\end{figure}

To evaluate the contributions of each component in AdaptiveLog, we conduct the ablation experiments. Specifically, two components of the AdaptiveLog are evaluated: Bayesian inference and ECR strategy. In fact, without the ECR strategy, the LLM analyzes hard samples directly using the standard ICL paradigm, where the LLM retrieves similar examples from the training set without the explicit reasoning process, the result has been given in Tables \ref{tab:main_exp1} - \ref{tab:main_exp3} and analyzed in RQ1. Obviously, the ECR strategy obtains a remarkable improvement compared to ICL for analyzing complex logs, validating the effectiveness of the ECR strategy. In addition, to further analyze the impact of the reasoning process in the case, we remove the \textit{Reason} part of the error-prone case, and the experimental results are shown in Table \ref{tab:ab-ecr}. The results indicate that the lack of explicit guidance from the reasoning process diminishes the performance of ECR strategy, affirming that potential pitfalls susceptible to errors help to remind the LLM to avoid similar mistakes.

Without Bayesian inference, the prior probability $p(C)$ will be used to estimate the uncertainty of the SLM. In addition, given the input log, considering that the selection of the LLM depends on the uncertainty of the SLM, we also compare it with the most straightforward routing strategy to further verify the Bayesian inference's effectiveness. Specifically, we train a binary classifier as the routing model to classify the input logs, the result True means that this log needs to query the LLM, otherwise directly outputs the result of the SLM. 
The training samples are derived from the results of the SLM on the validation set, the error samples of the SLM are marked as True otherwise as False. 
Finally, the results are shown in Fig. \ref{fig:Ablation}.

\begin{figure}[]
    \centering
    \subfloat[Huawei-Switches]{\label{fig: Hw-Switches}\includegraphics[width=0.4\linewidth]{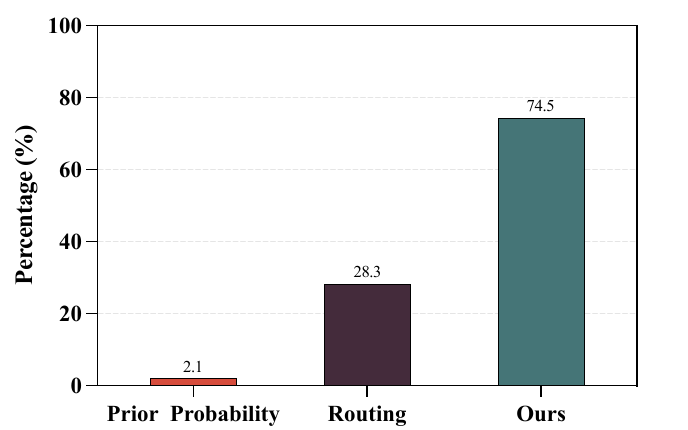}}
    \subfloat[Huawei-Routers]{\label{fig: system type zero}\includegraphics[width=0.4\linewidth]{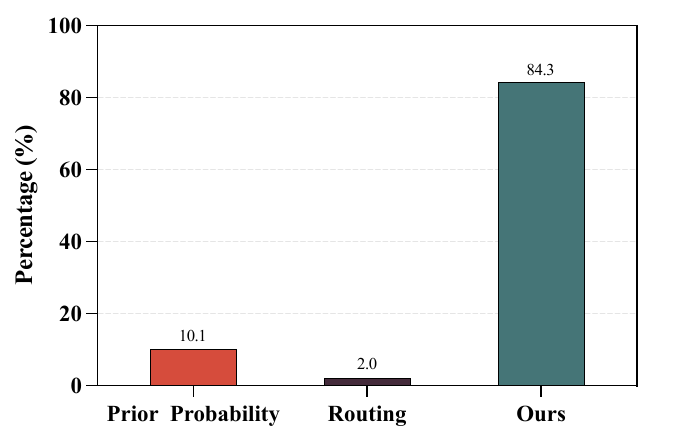}}
    \caption{Percentage of hard samples selected correctly.}
    \label{fig:Percentage}
\end{figure}

From the results, we observe that the model's performance drops significantly when Bayesian inference is not utilized, which validates that estimating model uncertainty through Bayesian inference can effectively filter out samples that exceed the capabilities of the SLM. Compared with the routing strategy, we find it difficult for the routing model to classify the input logs effectively. And even appears to be weaker than the SLM's (BERT) performance,  which indicates that the trained routing model cannot effectively find the mechanism for selecting hard samples from the input logs. To analyze the effectiveness of different strategies more intuitively, we also calculate the error sample percentage where the SLM is genuinely incorrect among the hard samples selected through different strategies in Fig. \ref{fig:Percentage}. Ideally, the selected hard samples should all be error samples of the SLM. We find that hard samples based on Bayesian inference account for the highest percentage of incorrect samples of the SLM and it has a significant advantage over prior probability and routing strategy in selecting hard samples, which is a prerequisite for efficiently exploiting the capabilities of LLM.

\subsubsection{Error-Prone Case Quality Ablation Studies}

\begin{table}[]
\centering
\caption{Results of the impact of error-prone case quality on ECR.}
\label{tab:quality}
\resizebox{0.55\columnwidth}{!}{%
\begin{tabular}{l|cc}
\toprule[1.5pt]
\multirow{2}{*}{}          & \multicolumn{2}{c}{LDSM (Accuracy / Weighted-F1)}     \\ \cmidrule{2-3} 
                           & \multicolumn{1}{c|}{Huawei-Switches} & Huawei-Routers \\ \midrule
Ratio of inconsistency      & \multicolumn{1}{c|}{1.3\% (2 / 155)}   & 2.5\% (5 / 196)  \\ \midrule
ECR with all Cases         & \multicolumn{1}{c|}{91.57 / 91.55}   & 92.50 / 92.53  \\
ECR w/o inconsistent cases & \multicolumn{1}{c|}{91.57 / 91.55}   & 92.50 / 92.53  \\ \bottomrule[1.5pt]
\end{tabular}%
}
\end{table}
A main component of AdaptiveLog is the error-prone case database, which is the backbone of the ECR strategy. To construct these cases, we leverage ChatGPT to analyze and explain the error-prone cases. Considering the quality of the error-prone cases is reliant on the quality of the ChatGPT, which will affect the ECR performance, we evaluate the quality of the error-prone cases on the performance of AdaptiveLog.

Firstly, we manually verified 20 cases from each dataset and found no significant errors. Secondly, to automatically assess the quality of error-prone cases, following self-consistency strategy \cite{wang2022self}, we increase the \textit{temperature} of ChatGPT (not zero) and re-generate two additional results, then calculate their similarity to the original results with \textit{bge-large-en-v1.5}. If the similarity is below 0.95, it is considered as the low-quality error-prone case. As shown in Table \ref{tab:quality}, it can be found that the proportion of low-quality error-prone cases is extremely low. Thirdly, to mitigate the impact of these low-quality error-prone cases, we remove these cases from the database and re-experiment with this database. As shown in Table \ref{tab:quality}, it can be seen that the removal of these low-quality error-prone cases have no effect on the results of the ECR strategy due to the extremely low proportion. This suggests that the error-prone cases constructed based on ChatGPT are nearly error-free, even if there are low-quality cases, we can mitigate the impact of low-quality cases through the self-consistency strategy.


\section{Discussion}
\subsection{Practicality}

\begin{figure}[]
    \centering
    \subfloat[Results on six different log analysis tasks]{\label{fig: aaaa}\includegraphics[width=0.5\linewidth]{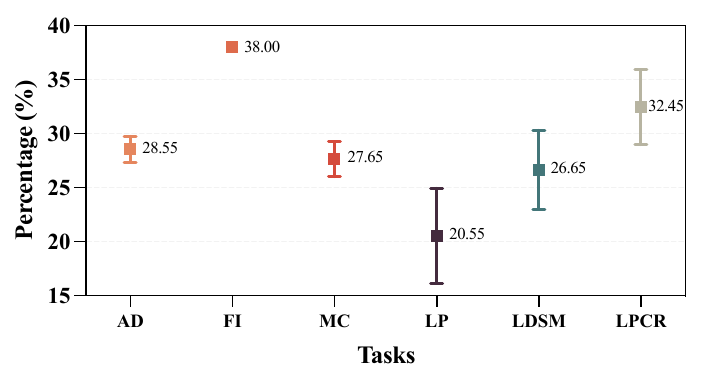}}
    \subfloat[Results in the low-resource and transfer scenario]{\label{fig: bbbb}\includegraphics[width=0.5\linewidth]{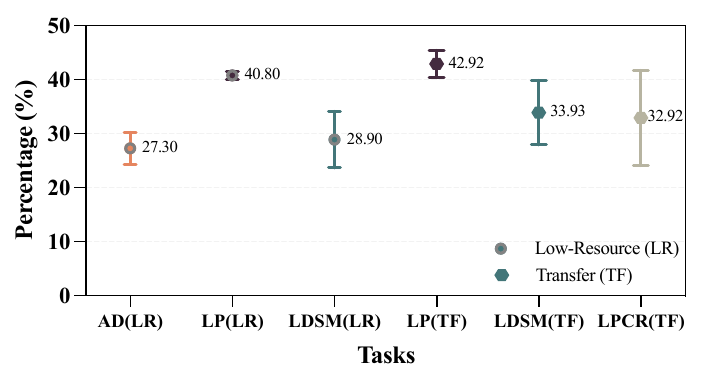}}
    \caption{Proportion of LLMs invoked by different tasks, where we count the proportion of hard samples for all datasets on each log analysis task.}
    \label{fig:box_cost}
\end{figure}

AdaptiveLog is designed to perform automated log analysis in production systems by collaborating with the SLM and the LLM. 
The framework aims to enhance log analysis performance while minimizing the cost associated with querying the LLM.
By estimating the uncertainty of the SLM and selectively invoking the LLM for uncertain results, AdaptiveLog effectively reduces the overall querying expense, optimizing performance without incurring significant LLM costs. 
Statistical analysis conducted on all six log analysis tasks under RQ1, as depicted in Fig. \ref{fig: aaaa}, reveals that on average, only 27\% of each dataset necessitates LLM querying. This approach results in a remarkable 73\% cost saving compared to analyzing all samples with the LLM, leading to state-of-the-art results.
Furthermore, an examination of the proportion of LLM queries in low-resource and transfer learning scenarios, illustrated in Fig. \ref{fig: bbbb}, indicates an increased need for LLM interventions in these challenging settings compared to the standard experimental setup of RQ1. This trend suggests the LLM's superior advantages in compensating for SLM limitations, particularly in more challenging scenarios. Despite the rise in LLM query percentages, AdaptiveLog still manages to save 65\% of the LLM overhead, showcasing its efficiency and adaptability across diverse operational contexts.

In practical applications, the construction of an error-prone case database plays a crucial role in enhancing the LLM's reasoning capabilities. The analysis of case database construction costs, detailed in Table \ref{tab:cost_database}, demonstrates that the expense and time costs associated with this process are minimal. Leveraging a sample size of 100 error-prone cases, the effectiveness of ECR is validated. To streamline case construction efforts, we suggest manually labeling high-uncertainty samples of the SLM to create error-prone cases, leveraging the tendency for high-uncertainty samples to be inaccurately predicted, thus reducing labeling efforts. Additionally, considering that constructing error-prone cases also requires ground truth, to further alleviate the labeling cost, we explore the performance of different numbers of case database on the ECR strategy.
Specifically, we employ the Determinantal Point Process (DPP) \cite{chen2018fast} algorithm to sample according to different proportions of error samples and ensure the diversity of samples. 
As shown in Table \ref{tab:error sample percent},  all metrics improve with an increased number of error samples. Even with only 60\% of error samples, the ECR strategy is still effectively superior to ICL. Considering the substantial increase in manual labeling efforts with more ground-truth datasets in practice, our results show that it is not necessary to construct all error samples, ECR is still effective.

\begin{table*}[]
\caption{The size and cost of all error-prone databases.}
\label{tab:cost_database}
\tabcolsep=0.07cm
\resizebox{\textwidth}{!}{%
\begin{tabular}{l|cc|c|cc|cc|cc|c}
\toprule[1.5pt]
Task              & \multicolumn{2}{c|}{Anomaly Detection} & FI        & \multicolumn{2}{c|}{Module Classification} & \multicolumn{2}{c|}{Level Prediction} & \multicolumn{2}{c|}{LDSM}      & LPCR           \\ \midrule
Dastsets          & BGL            & ThunderBird           & Openstack & Cisco                & Huawei              & Cisco             & Huawei            & Cisco          & Huawei        & Huawei         \\ \midrule
Num               & 38             & 189                   & 20        & 72 (39+33)           & 46 (15+31)          & 133 (61+72)       & 93 (64+29)        & 662 (373+289)  & 351 (155+196) & 498 (262+236)  \\
Total Prices (\$) & 0.2            & 0.8                   & 0.1       & 0.4 (0.2+0.2)        & 0.3 (0.1+0.2)       & 0.6 (0.3+0.3)     & 0.5 (0.3+0.2)     & 2.7 (1.5+1.2)  & 1.4 (0.6+0.8) & 2.1 (1.1+1.0)  \\
Total Time (s)    & 86             & 427                   & 45        & 162 (88+74)          & 104 (34+70)         & 300 (138+162)     & 210 (145+65)      & 1494 (842+652) & 792 (150+442) & 1124 (591+533) \\ \bottomrule[1.5pt]
\end{tabular}%
}
\end{table*}

\begin{figure}[]
    \centering
    \subfloat[Results only on hard Samples (HW-Switches)]{\label{fig: Hw-Switches}\includegraphics[width=0.45\linewidth]{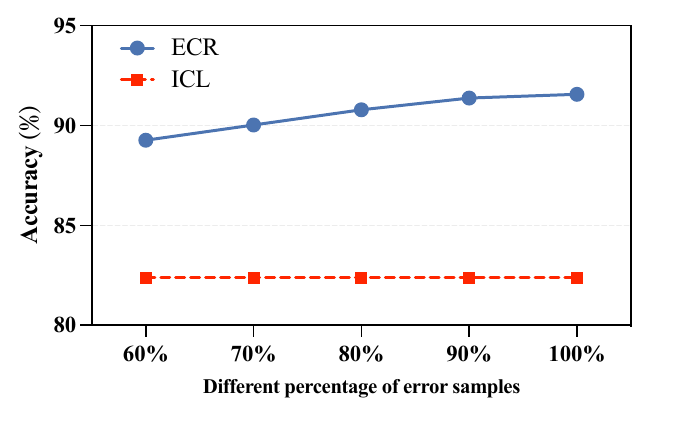}}
    \subfloat[Results only on hard Samples (HW-Routers)]{\label{fig: system type zero}\includegraphics[width=0.45\linewidth]{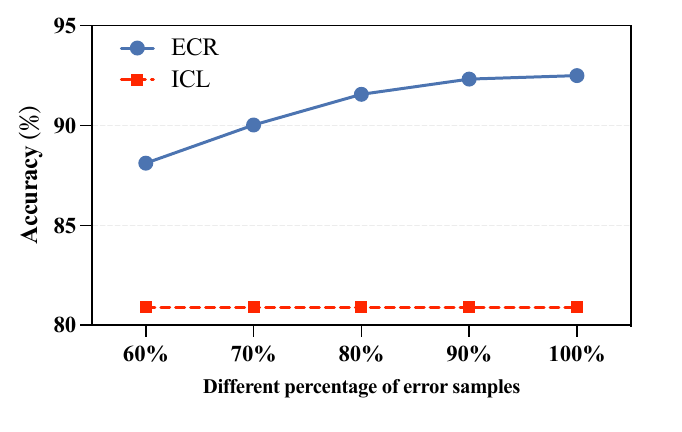}}
    \caption{ECR results based on different proportions of error-prone cases on the LDSM.}
    \label{fig:error_case_proportions}
\end{figure}

\begin{figure}[]
    \centering
    \subfloat[Hw-Switches Correct Samples]{\label{fig: Hw-Switches}\includegraphics[width=0.4\linewidth]{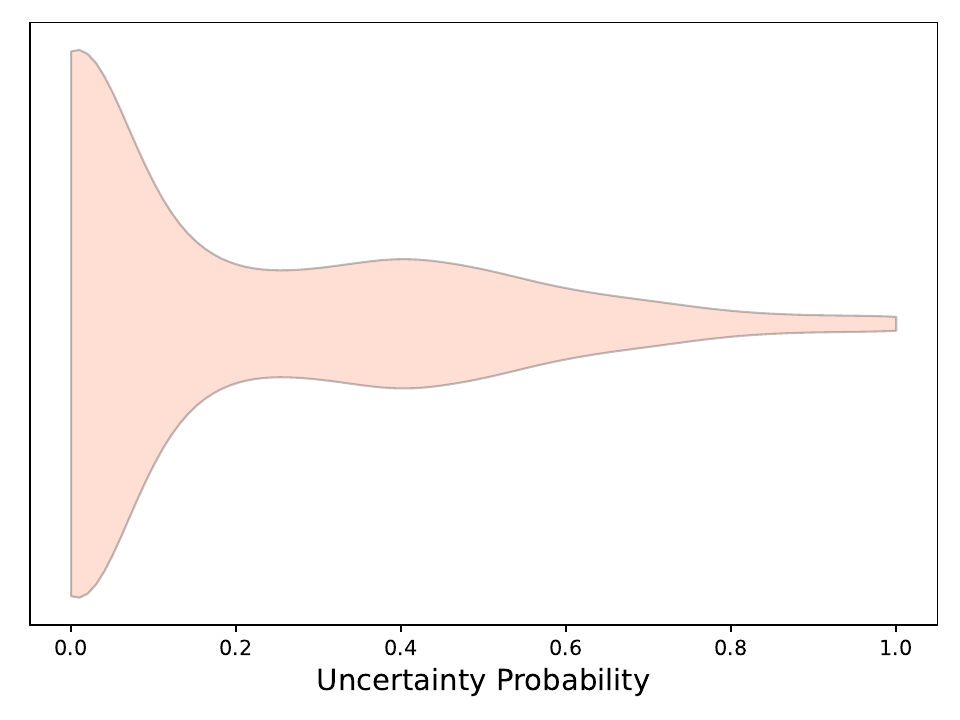}}
    \subfloat[Hw-Switches Error Samples]{\label{fig: system type zero}\includegraphics[width=0.4\linewidth]{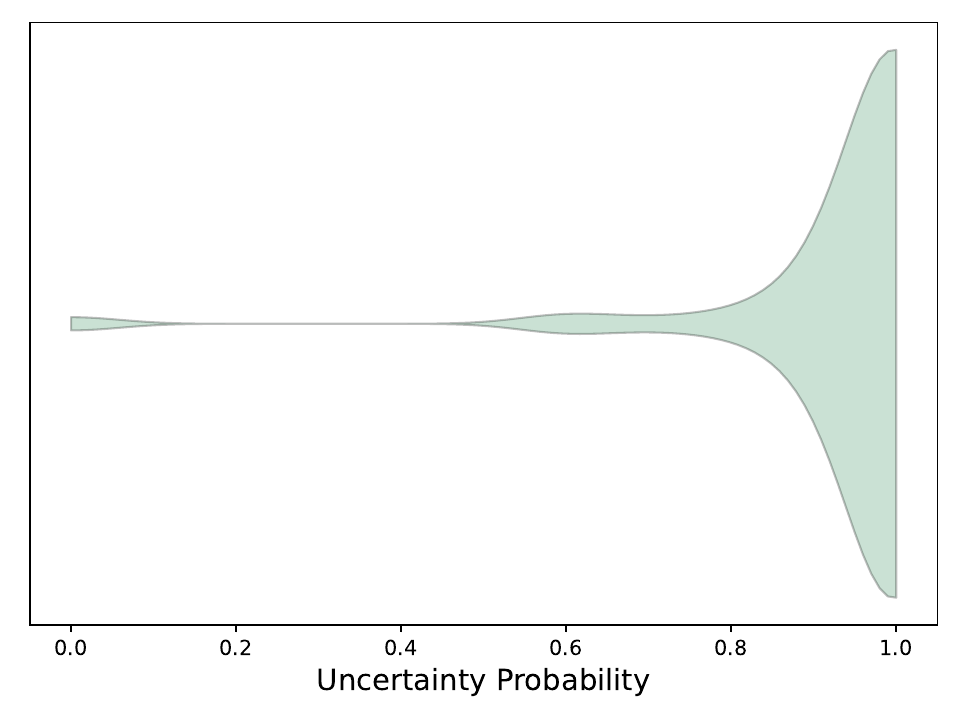}}
    
    \subfloat[Hw-Routers Correct Samples]{\label{fig: Hw-Switches}\includegraphics[width=0.4\linewidth]{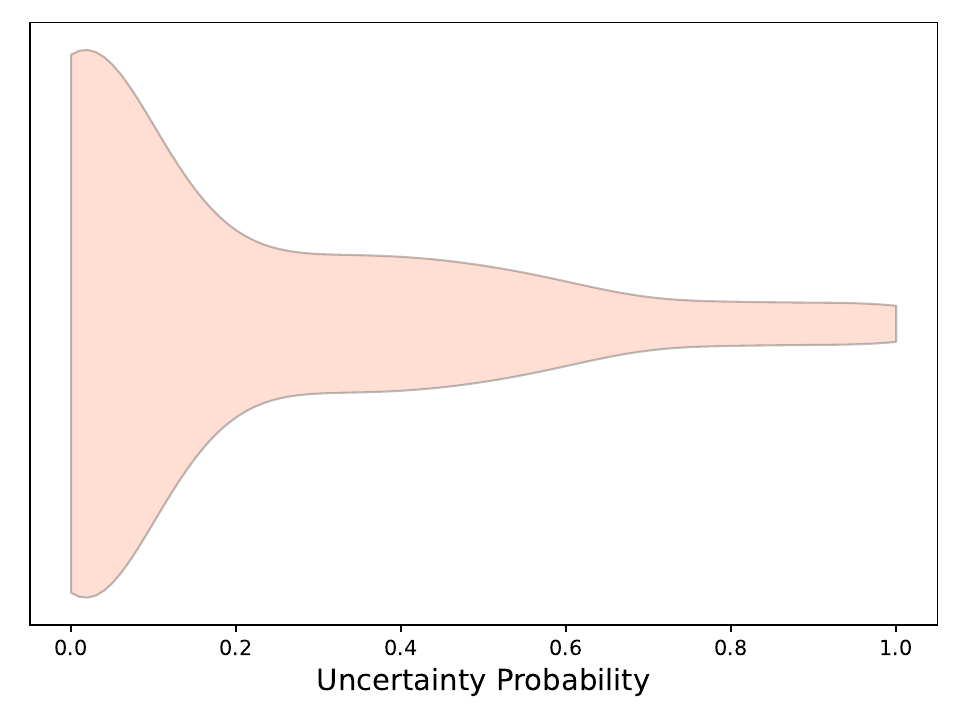}}
    \subfloat[Hw-Routers Error Samples]{\label{fig: system type zero}\includegraphics[width=0.4\linewidth]{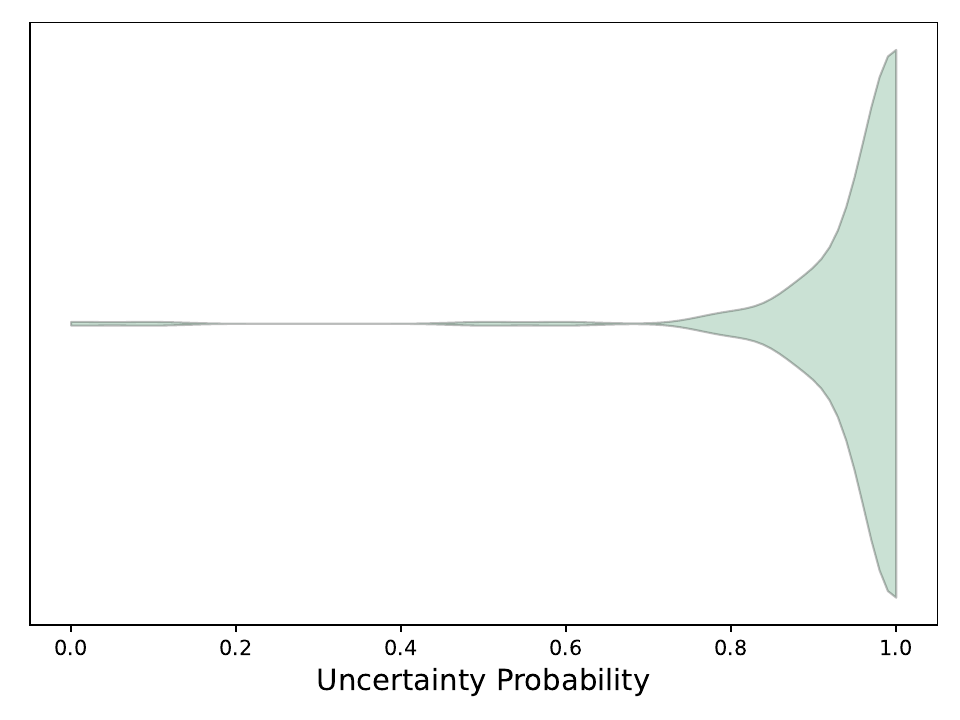}}
    \caption{Uncertainty distribution chart on LDSM task.}
    \label{fig:distribution}
\end{figure}

It is worth noting that AdaptiveLog is compatible with different pre-trained language models, such as KnowLog. According to the experimental results in RQ4, AdaptiveLog, when integrated with other SLMs, can also achieve higher performance than state-of-the-art log analysis models. This depends on the fact that Bayesian inference can effectively estimate the uncertainty of SLMs, and low uncertainty is often beyond the capacity of models. To delve deeper into the relationship between model uncertainty and performance, we plot the distribution of uncertainty between correct and error samples on the LDSM task. As shown in Fig. \ref{fig:distribution}, the distribution reveals a high correlation between uncertainty and performance. From the distribution, it can be seen that the probability of uncertainty for error samples that are incorrectly predicted by the SLM tends to cluster between 0.8 and 1.0,  while the uncertainty for correct samples mainly falls between 0 and 0.4. This observation implies that models are significantly prone to errors when their uncertainty probability is low. Consequently, the LLM is required to analyze only those samples surpassing the SLM's capabilities, thereby diminishing the overhead of the LLM.
In conclusion, the practicality of AdaptiveLog in real-world systems is evident, with specialized development and deployment, the log analysis process will be significantly optimized.

\begin{figure}[]
\centering
   \includegraphics[width=0.8\textwidth]{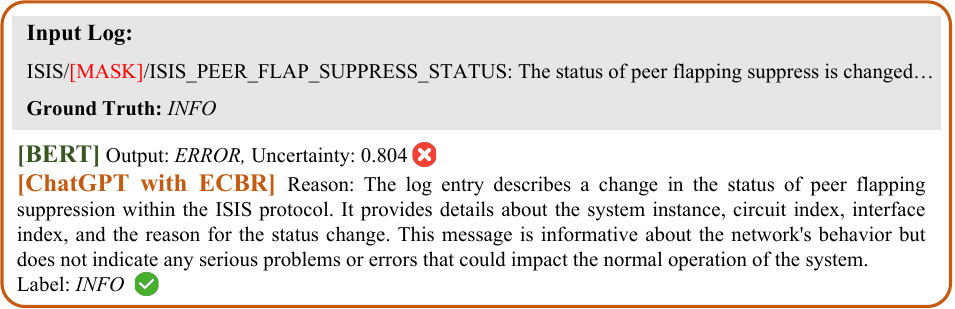}
   \caption{Qualitative example on the Level Prediction task.}
   \label{fig:qualitative case}
\end{figure}

\begin{table}[]
\centering
\caption{Qualitative analysis of LLM interventions on the LDSM task.}
\label{tab:qualitative}
\resizebox{\columnwidth}{!}{%
\begin{tabular}{l|ccc|ccc}
\toprule[1.5pt]
\multirow{2}{*}{Datasets} & \multicolumn{3}{c|}{LLM Incorrect Intervention}                                                                                                                                                                                                                  & \multicolumn{3}{c}{LLM Correct Intervention}                                                                                                                                                                                                                     \\ \cmidrule{2-7} 
                          & \multicolumn{1}{c|}{\begin{tabular}[c]{@{}c@{}}SLM correct predictions \\ (\#Num)\end{tabular}} & \multicolumn{1}{c|}{\begin{tabular}[c]{@{}c@{}}LLM incorrect predictions \\ (\#Num)\end{tabular}} & \begin{tabular}[c]{@{}c@{}}Percentage \\ (\%)\end{tabular} & \multicolumn{1}{c|}{\begin{tabular}[c]{@{}c@{}}SLM incorrect predictions \\ (\#Num)\end{tabular}} & \multicolumn{1}{c|}{\begin{tabular}[c]{@{}c@{}}LLM correct predictions \\ (\#Num)\end{tabular}} & \begin{tabular}[c]{@{}c@{}}Percentage \\ (\%)\end{tabular} \\ \midrule
Huawei-Switches               & \multicolumn{1}{c|}{417}                                                                        & \multicolumn{1}{c|}{38}                                                                           & 9.11                                                       & \multicolumn{1}{c|}{105}                                                                          & \multicolumn{1}{c|}{99}                                                                         & 94.28                                                      \\
Huawei-Routers                & \multicolumn{1}{c|}{380}                                                                        & \multicolumn{1}{c|}{30}                                                                           & 7.89                                                       & \multicolumn{1}{c|}{154}                                                                          & \multicolumn{1}{c|}{144}                                                                        & 93.50                                                      \\ \bottomrule[1.5pt]
\end{tabular}%
}
\end{table}

\subsection{Qualitative Analysis}
In this section, we conduct a qualitative analysis on the Level Prediction task, presented in Fig. \ref{fig:qualitative case}. The analysis focused on the behavior of collaboration between BERT and ChatGPT. 
Initially, BERT is employed to provide a prediction along with an associated probability of uncertainty. This uncertainty estimation determines the invocation of the ChatGPT, as outlined in Equation \ref{select}. 
Due to the limited capacity of BERT, it exhibits higher uncertainty, and the selection is made to query ChatGPT in this case. Upon querying ChatGPT, we observe that it conducts a thorough analysis of the log content by leveraging its extensive knowledge and relevant case experiences. Subsequently, after a thorough evaluation of the log content and determining its lack of relevance to the identified risk, ChatGPT gives a reasonable prediction.
Hence, the integration of ChatGPT into the analysis workflow yielded notable benefits. 
The insufficient capabilities of the SLM underscores
the LLM's value in enhancing the analysis process, its contribution significantly improves the accuracy and reliability of log analysis tasks.

In addition, given AdaptiveLog's uncertainty-based LLM invocation mechanism, 
a potential concern arises regarding the possibility of LLM interventions overriding correct predictions made by the SLM due to uncertainty estimation. To delve into the impact of LLM intervention on uncertain samples, we perform a qualitative analysis on the LDSM task.  The results presented in Table \ref{tab:qualitative} reveal that the LLM makes fewer than 10\% incorrect interventions, with over 90\% being correct interventions. This outcome signifies a positive effect of LLM intervention, highlighting its ability to effectively complement the SLM's capabilities and compensate for its limitations through collaborative decision-making.
This collaborative framework not only enhances performance but also underscores the significance of leveraging diverse model strengths for comprehensive log analysis, reinforcing the framework's efficacy and potential in real-world log analysis scenarios.

\subsection{Threat to Validity}
We identified the following major threats to validity:

\textbf{Data Leakage:} One significant threat to the validity of our study is the potential for data leakage in LLMs. These models are trained on the extensive corpus, raising concerns that they might inadvertently memorize specific patterns or results from the training data, rather than genuinely performing inference. However, our experimental results suggest that this risk is minimal. 
On the one hand, from the results of different prompt strategies, LLMs are significantly affected by the prompt. Especially the random examples can seriously affect the performance of LLMs, indicating that the model is not merely recalling memorized data but is instead leveraging contextual information to enhance its reasoning. 
On the other hand, the majority of our experiments utilize the \textit{gpt-3.5-turbo-16k} model, apart from utilizing the software system logs collected from Loghub, we also collect network device logs from the latest public documentation of different network device vendors for our experiments. It is noteworthy that this model version ceased receiving updates before the release of the latest public documentation. As a result, it is highly improbable that these network device logs were included in the training data for the version of the model we used. Thus, we consider the likelihood of data leakage affecting our experimental results to be negligible.

\textbf{Randomness:} Randomness may affect the validity of results obtained from the LLM, influencing their performance in distinct ways. This threat manifests in two primary aspects within our study: (1) Randomness in LLM inference: The inherent randomness present in the inference process of the LLM can lead to variations in their outputs for the same input logs. To address this concern, we set the \textit{model temperature} to 0. This adjustment ensures that the LLM consistently generates the same outputs for identical input logs, thereby minimizing the impact of inherent randomness on our results.
(2) Randomness in the selection of error cases in RQ4: Random selection of prompt demonstrations introduces variability in the data, this may skew the model's understanding and decision-making processes. To mitigate the threat, we select demonstrations three times to conduct experiments. By averaging the results obtained from the LLM, we derive a final result that reflects a more stable and reliable outcome. 
\section{Conclusion}
In conclusion, this paper introduces AdaptiveLog, a pioneering log analysis framework that synergizes an LLM and an SLM, effectively balancing performance and inference costs to achieve optimal results while significantly reducing the computational expenses associated with the LLM. AdaptiveLog's innovative approach lies in its adaptive selection strategy, which triggers LLM queries based on the uncertainty estimation of the SLM, ensuring LLM utilization only in cases where the SLM's uncertainty is high.
Moreover, AdaptiveLog incorporates an Error-Case Reasoning Enhancement  (ECR) strategy to retrieve similar error-prone cases, bolstering the reasoning capabilities of the LLM in log analysis tasks. 
Extensive experimentation validates the effectiveness of AdaptiveLog, showcasing state-of-the-art results across various log analysis tasks while maintaining efficiency and cost-effectiveness.
We believe that AdaptiveLog can offer a compelling solution for practitioners and researchers in the log analysis field.

\section{Data Availability}
Our source
code, detailed experimental data, and results are available at \url{https://github.com/LeaperOvO/AdaptiveLog-review}.

\bibliographystyle{ACM-Reference-Format}
\bibliography{ref}

\end{document}